\documentclass[twocolumn,noshowpacs,preprintnumbers,amsmath,amssymb,pre]{revtex4}
\usepackage{graphicx}
\usepackage{bm}

\newcommand{\EQ}[1] {\begin{equation}#1\end{equation}}
\newcommand{\EQA}[1] {\begin{eqnarray}#1\end{eqnarray}}
\newcommand{\PD} {\partial}
\newcommand{\EQAMK} {&\!\!\!}
\newcommand{\EQAEQ} {\EQAMK=\EQAMK}

\newcommand{\SE} {s}
\newcommand{\CA} {\check{A}}

\begin{document}

\title{
Turbulent Relative Dispersion in Two-Dimensional Free Convection
Turbulence
}

\author{Takeshi OGASAWARA}
\email{ogasawara@kyoryu.scphys.kyoto-u.ac.jp}
\author{Sadayoshi TOH}
\email{toh@scphys.kyoto-u.ac.jp}
\affiliation{
Division of Physics and Astronomy, Graduate School of Science, Kyoto
University, Kyoto 606-8502, Japan}

\begin{abstract}
 The relative dispersion process in two-dimensional free convection
 turbulence is investigated by direct numerical simulation.
 In the inertial range, the growth of relative
 separation, $r$, is expected
 as $\langle r^2(t)\rangle\propto t^5$ according to the Bolgiano--Obukhov
 scaling.
 The result supporting the scaling is obtained with exit-time
 statistics.
 Detailed investigation of exit-time PDF shows
 that the PDF is divided into two regions,
 the Region-I and -II,
 reflecting two types of separating processes:
 persistent expansion and random
 transitions between expansion and compression of relative separation.
 This is consistent with the physical picture of the
 self-similar telegraph model.
 In addition, 
 a method for estimating the parameters of the model are presented.
 Comparing two turbulence cases, two-dimensional free convection and
 inverse cascade turbulence,
 the relation between the drift term of the model and nature of coherent
 structures is discussed.
\end{abstract}

\maketitle

 \section{Introduction}
 Relative dispersion of passive particles in turbulent flows is one of
 the fundamental problems in turbulence
 research.
 It characterizes the transport and mixing properties of turbulence and
 is important from both theoretical and practical points of view.
 Reflecting universal behavior of turbulent fluctuations, relative
 dispersion also has some universal properties because of its
 locality-in-scale nature \cite{B1950,MY1975,FV1998}.
 In particular, the dispersion process exhibits anomalous dispersion
 in the inertial range.
 This is first observed by Richardson (1926) \cite{R1926},
 and since then, a number of theoretical, experimental, and numerical
 investigations have been devoted to
 understand and
 model relative dispersion process
 \cite{JPT1999,OM2000,IK2002,BS2002a,BS2002b,GV2004,RE2005,S2001}.
 However, comprehensive understanding
 has
 not been obtained yet.

 Recently, a few works focusing on underlying mechanism of
 the anomalous dispersion were reported.
 In the inertial range, the mean free path, $l(r)$, the mean length 
 for persistent expansion of relative separation
 without changing its moving direction, is an order of relative
 separation itself, $r$: $l(r)=P_s r$ \cite{S1999,SKB2000}.
 Here, $P_s$ is a non-dimensional constant called the persistent
 parameter.
 In two-dimensional inverse cascade turbulence (2D-IC), $P_s$ is
 estimated as $0.87$ \cite{BS2002a}. This means separating motions are
 not purely diffusive but composed of an appreciable amount of
 persistent motions.
 In addition, it was also reported that there is a relation between 
 stagnation-point structures and Richardson's law \cite{DV2003},
 and that dispersion process is described by persistent streamline
 topology \cite{GV2004}.
 From these results, it is expected that coherence of turbulent field,
 which must share its origin with fine coherent vortical structures such
 as worms in three-dimensional Navier-Stokes (3D-NS)
 turbulence, has a significant role in turbulent relative dispersion.
 
 The correlations in turbulence are characterized in scale-space due to
 their self-similarity,
 and  are not made disappear by coarse
 graining in real space.
 Reflecting these correlations,
 relative separation moves persistently
 to some extent,
 so that,
 unlike the Brownian motion,
 relative separation
 process
 should not be
 described only by random
 collision motions
 even as an approximation\cite{S1999}.
 In other words, the characteristic length cannot be defined
 because the mean free path, $l(r)$, varies depending on
 the spatial scale.
 Whereas there are several experiments and numerical simulations
 of which results are rather close to the prediction of Richardson's
 diffusion equation \cite{JPT1999,BS2002a,GV2004}
 that closely relates to random collision motions.
 Thus, these results raise a question;
 how are the effects of persistent motions wiped out?
 In the previous paper \cite{OT2006b},
 we have introduced a self-similar telegraph
 model of turbulent relative dispersion, and showed that
 the separation PDF
 can be close to the prediction of Richardson's diffusion equation
 for slowly separating particle pairs
 even in the presence of persistent motions.
 
 In the present paper,
 we check the consistency of the physical picture of the self-similar
 telegraph model
 by carrying
 out direct numerical simulations (DNS) of
 2D free convection (2D-FC) turbulence
 instead of
 3D-NS
 turbulence.
 This is
 because
 (i) DNS of 3D-NS turbulence requires extremely large
 computer resources, so that it is difficult to track particles for a
 long time, which is necessary to investigate dynamical properties of
 relative dispersion process,
 and (ii)
 2D-FC turbulence has both statistical and dynamical
 characteristics similar to those of 3D-NS turbulence
 (see Appendix A)
 \cite{TS1994,TM2003}.
 Among them, the existence of coherent
 structures, which are approximated 
 by the Burgers T-Vortex layer, is notable.
 This is a crucial difference from 
 coherent structures in 2D-IC
 turbulence that are nested cat's eye vortices
 \cite{FV1998,GV2004}.
 Therefore, comparing the
 results of the 2D-FC case with those of the 2D-IC case,
 we can investigate the effects of coherent structure on
 turbulent relative dispersion.
 This comparison gives
 a physical meaning of the drift term of the self-similar telegraph model.

 The inertial range achieved by our DNS is not so wide that 
 the relative motions of particle-pairs in the dissipation, 
 the inertial, and the energy containing scales are
 not 
 sufficiently 
 resolved  by usual
 fixed time statistics. 
 In order to investigate the scaling natures of relative dispersion 
 in such  a narrow and limited inertial range,
 we utilize exit-time statistics introduced 
 into research of turbulent relative dispersion by
 Boffetta and Sokolov
 (see Appendix B) \cite{ABCCV1997,BCCA1999,BS2002a}.
 By detailed investigation of the PDF of exit-time,
 we show that the PDF is divided into two region, the Region-I and -II,
 corresponding to persistent expansion and random transition between
 expansion and compression of relative separation, respectively.
 This result agrees with the picture of the self-similar telegraph model.
 In addition, we provide a method for estimation of the parameters of
 the self-similar telegraph model by using exit-time PDF.

 In the following sections, first we provide a brief review of the
 self-similar telegraph model in \S 2.
 Section 3 presents a summary of the numerical scheme and parameters
 of our DNS of 2D-FC turbulence.
 Some of the results by the fixed-time and
 the exit-time statistics are provided and discussed in \S 4.
 Concluding remarks are made in \S 5.
 In addition,
 some properties of 2D-FC turbulence and exit-time statistics are
 presented in Appendix A and B, respectively.

 \section{Self-similar telegraph model and Palm's equation}
 In the previous paper \cite{OT2006b},
 we have introduced a self-similar telegraph model
 for turbulent relative dispersion,
 which is a model describing the evolution of
 the PDF of relative separation of particle pairs
 in the inertial range.
 The relative separation of two particles, $r(t)$ is defined as follows:
 \begin{equation}
  r(t)=|\bm{x}_1(t)-\bm{x}_2(t)|,  
 \end{equation}
 where $\bm{x}_1(t)$ and $\bm{x}_2(t)$ are the Lagrangian positions 
 of the particles.
 The model
 is based on Sokolov's model \cite{S1999} and
 consists of persistent expansion and compression of relative
 separation, $r$,
 according to the relative velocity,
 $v(r)=\CA r^{1-\SE}$,
 where $\CA$ and $\SE$ are a
 dimensional
 constant and a scaling
 exponent, respectively: $\SE=2/3$ for Kolmogorov scaling and 
 $\SE=2/5$ for Bolgiano-Obukhov scaling.
 The transition rate from expansion to compression and
 that from compression to expansion are given by
 $\lambda^+/T_c(r)$ and $\lambda^-/T_c(r)$, respectively.
 Here $T_c(r)$ is a characteristic time scale at a spatial scale $r$,
 $T_c(r)=\CA^{-1}r^\SE$,
 and $\lambda^\pm$ are the inverses of persistent
 parameters, $P_+$ for expansion and $P_-$ for compression.
 Introducing $\lambda=\lambda^+ + \lambda^-$ and
 $\delta=\lambda^+ - \lambda^-$,
 the evolution equation of the separation PDF, $P(r,t)$, 
 is derived as follows:
 \EQA{
 \frac{T_c(r)}{\lambda}
 \frac{\PD^2 P}{\PD t^2}
 +
 \frac{\PD P}{\PD t}
 \nonumber
 =
 \frac{\PD}{\PD r}
 \left[
 D(r)r^{d-1}\frac{\PD}{\PD r}\left(\frac{P}{r^{d-1}}\right)
 \right]
 \\
 +\sigma\frac{\PD}{\PD r}
 \left[v(r)P\right]
 ,
 \label{eq:T-Model}
 }
 where
 $D(r)$ is Richardson's diffusion coefficient,
 $\CA\lambda^{-1}r^{2-\SE}$,
 and
 $\sigma=(d-2\SE+\delta)\lambda^{-1}\equiv\tilde{\sigma}\lambda^{-1}$.
 The parameters of the model are $\lambda$ and $\delta$:
 $\lambda$ represents the strength of persistency of moving directions,
 and $\delta$ does the difference in persistency between expansion and
 compression.
 Therefore, $\lambda$ and  $\delta$ 
 reflect the strength and structure of coherence of the flow, 
 respectively. 

 For slowly-separating particle pairs,
 i.e.,
 in the case of $r\ll\langle r^2\rangle^{1/2}$, the first term of the
 l.h.s.\ of Eq.\ (\ref{eq:T-Model}) can be neglected and
 the approximated equation is given by
 \EQ{
 \frac{\PD P}{\PD t}
 =
 \frac{\PD}{\PD r}
 \left[
 D(r)r^{d-1}\frac{\PD}{\PD r}\left(\frac{P}{r^{d-1}}\right)
 \right]
 +\sigma\frac{\PD}{\PD r}
 \left[v(r)P\right].
 \label{eq:Palm-eq}
 }
 This form of the equation was first derived by Palm \cite{MY1975} and
 is also the same as the diffusion equation of Goto-Vassilicos model
 \cite{GV2004}.
 We call Eq.\ (\ref{eq:Palm-eq}) Palm's equation.
 The similarity solution of Eq.\ (\ref{eq:Palm-eq}) is
 \EQ{
 P(r,t)
 =
 C_P t^{-\frac{1}{\SE}}
 \left(
 \frac{\lambda r^\SE}{\SE^2 \CA t}
 \right)^{\frac{d-\tilde{\sigma}-1}{\SE}}
 \exp
 \left(
 -\frac{1}{\SE^2}
 \frac{\lambda r^\SE}{\CA t}
 \right),
 \label{eq:Palm-Solution}
 }
 where $C_P$ is the normalization factor.

 Because the tail of the exit-time PDF, $P_E(T_E; r, \rho)$, 
 consists of slowly-separating particle pairs,
 it is calculated from Palm's equation (\ref{eq:Palm-eq})
 with
 the method used 
 by Boffetta and Sokolov \cite{BS2002a}.
 The asymptotic form of $P_E(T_E; r, \rho)$ is given by
 \EQA{
 P_E(T_E; r, \rho)
 \qquad\qquad\qquad\qquad
 \qquad\qquad\qquad\quad
 \nonumber
 \\
 \sim
 \exp\left(
 -\frac{1}{4}
 \frac{\SE j^2_{1-\frac{\delta}{\SE},1}}{2\SE-\delta}
 (1-\rho^{-\SE})
 \frac{T_E}{\langle T_E(r;\rho)\rangle}
 \right),
 \label{eq:Palm-etime}
 }
 where $j_{\nu,n}$ is the $n$-th zero of the $\nu$-th order Bessel
 function and $\langle T_E(r;\rho)\rangle$ is the mean exit-time from $r$
 to $\rho r$:
 \EQA{
 \langle T_E(r;\rho)\rangle
 =
 \frac{\lambda}{\CA}
 \frac{1}{\SE(2\SE-\delta)}
 \left(
 \rho^\SE-1
 \right)r^\SE.
 \label{eq:avetime-palm}
 }
 Note that the mean exit-time calculated from the self-similar telegraph
 model, Eq.\ (\ref{eq:T-Model}), is the same as Eq.\
 (\ref{eq:avetime-palm}).
 This is because the mean exit-time is calculated from a steady
 solution of the equation \cite{BS2002a},
 and the solution of Eq.\ (\ref{eq:T-Model}) is
 the same as that of Eq.\ (\ref{eq:Palm-eq}).
 By comparing the tail of the exit-time PDF obtained by DNS and 
 Eq.\ (\ref{eq:Palm-etime}), we can estimate the value of $\delta$.
 
 The last term of the r.h.s.\ of Eq.\ (\ref{eq:T-Model}) is a drift term consistent with
 the scaling law, and the direction of the drift is determined 
 by $\tilde{\sigma}$.
 The parameter $\tilde{\sigma}$ consists of two parts, the ``scaling-determined''
 one, $d-2\SE$, and the ``dynamics-determined'' one, $\delta$.
 In order for Eq.\ (\ref{eq:Palm-eq}) to recover Richardson's equation,
 the drift term has to disappear, which means $\delta=2\SE-d\equiv\delta_0$.
 We call this case the Richardson case or the zero-drift case,
 where the parameters of the model reduce to one, 
 $\lambda^+$;
 $\lambda^-$ is determined by the relation
 $\lambda^-=\lambda^+-\delta_0\equiv\lambda^-_0$.

 \section{Numerical Simulation}
 \begin{figure}
  \begin{center}
  \includegraphics[width=8.8cm]{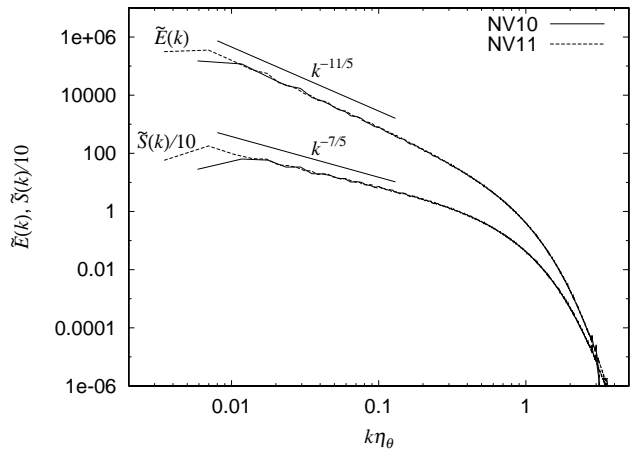}
  \end{center}
  \caption{\label{fig:Spectrum}
  Rescaled energy and entropy spectrum $\tilde{E}(k)$ and
  $\tilde{S}(k)$ obtained by DNS
  at resolution $N=1024$ (NV10) and $N=2048$ (NV11).
  The straight lines refer to the Bolgiano-Obukhov scaling:
  $\tilde{E}(k\eta_\theta)\equiv \eta_\theta^{-3} \tau_\theta^{-2}E(k)$,
  $\tilde{S}(k\eta_\theta)\equiv \eta_\theta^{-1} \epsilon_\theta^{-1}
  \tau_\theta^{-1}S(k)$.
  }
 \end{figure}

 \begin{table}
  \caption{\label{table:dnsparms}
  Parameters used in the present DNS.
  }
  \begin{ruledtabular}
   \begin{tabular}{cccccc}
    label & $N$ & $\nu$ & $f_0$ & $d_0$ & $k_d$ \\
    \hline
    NV10 & $1024$ & $1\times10^{-4}$ & $5\times10^{-2}$ &
    $1\times10^{-1}$ & 3 \\
    NV11 & $2048$ & $5\times10^{-5}$ & $5\times10^{-2}$ &
    $1.5\times10^{-1}$ & 10 \\
   \end{tabular}
  \end{ruledtabular}
 \end{table}

 \begin{table*}
  \caption{\label{table:char_quantities}
  Characteristic quantities of generated turbulent
  fields.
  $\Delta x$ represents the grid size.
  }
  \begin{ruledtabular}
   \begin{tabular}{ccccccc}
    label & $\epsilon_\theta$ & $\eta_\theta$ & $\tau_\theta$ &
    $\lambda_x$ & $\lambda_y$ & $\text{Ra}_\lambda$ \\
    \hline
    NV10 & $7.11\times10^{-3}$ & $5.87\times10^{-3}$ ($0.956\Delta x$) &
    $3.44\times10^{-1}$ & $4.50\times10^{-2}$& $4.54\times10^{-2}$ &
    $2.50\times10^{3}$\\
    NV11 & $1.44\times10^{-2}$ & $3.49\times10^{-3}$ ($1.14\Delta x$) &
    $2.43\times10^{-1}$ & $3.01\times10^{-2}$ & $3.03\times10^{-2}$ &
    $3.99\times10^{3}$ \\
   \end{tabular}
  \end{ruledtabular}
 \end{table*}
 
 In this section, we explain the method of DNS used in
 the present work and show basic properties of turbulent fields produced by
 our simulation.

 We generate turbulent field by DNS of the vorticity equation with a
 large-scale friction term $F_d$ and the temperature equation with a
 large-scale forcing term $F_f$:
 \EQA{
 \nabla\cdot\bm{u}=0,
 \\
 \frac{\PD\omega}{\PD t}
 + (\bm{u}\cdot\nabla)\omega =
 \nu\triangle\omega + \alpha g
 \frac{\PD T}{\PD x} + F_d,
 \\
 \frac{\PD T}{\PD t}
 + (\bm{u}\cdot\nabla) T =
 \kappa\triangle T + F_f,
 }
 where $\omega$, $T$, $\bm{u}$, $\nu$, $\kappa$, $\alpha$, and $g$ represent
 the vorticity field, the temperature field,
 the velocity field,
 the kinematic viscosity,
 the thermal diffusivity, the thermal expansion coefficient, and the
 gravitational acceleration, respectively.
 The large-scale forcing term used here is
 \EQA{
 F_f(\bm{x}) = 4 f_0 \cos(2x) \cos(2y),
 }
 where $f_0$ is a constant.
 The large-scale friction term is written in the Fourier space as
 \EQA{
 \hat{F}_d(\bm{k}) =
 \begin{cases}
  - \frac{d_0}{|\bm{k}|^2} \hat{\omega}(\bm{k})
  &
  (0<|\bm{k}|\le k_d),
  \\
  0
  &
  (\text{otherwise}),
 \end{cases}
 }
 where $d_0$, $\bm{k}$, $\hat{F}_d(\bm{k})$, and $\hat{\omega}(\bm{k})$
 are a constant, the wave number vector, the Fourier mode of the
 friction term, and the Fourier mode of the vorticity
 field,
 respectively.
 Our DNS is performed on a $2\pi\times2\pi$ domain with the doubly periodic
 boundary conditions at resolutions
 $N^2$: $N=1024$ (NV10) and $2048$ (NV11).
 We employ a pseudo-spectral method
 for accurately calculating convolutions and
 spatial derivatives,
 and 4-th order Runge-Kutta method for time integration.
 Aliasing error is removed by adopting the phase-shift method (NV10) and
 the $3/2$ method (NV11).
 All results presented in this paper are obtained
 for
 statistically stationary and locally isotropic turbulence.
 We summarize 
 the parameters used in our simulation 
 in Table \ref{table:dnsparms} and
 characteristic quantities of generated turbulence
 in Table \ref{table:char_quantities}.

 Figure \ref{fig:Spectrum} shows the entropy and energy spectra obtained by
 our DNS.
 The spectra are rescaled with entropy dissipation scales and the
 entropy dissipation rate, and
 there is a region consistent with the Bolgiano-Obukhov scaling,
 Eqs.\ (\ref{eq:BO-Ek}) and (\ref{eq:BO-Sk}),
 (see Appendix A).
 We call this region the inertial range.
 Most investigations in the present paper are carried out in this range.

 In the velocity field generated by DNS,
 we track a number of particle pairs ($1\times10^6$ pairs)
 according to the advection equation:
 \EQA{
 \frac{\PD}{\PD t} \bm{x}_i(t) = \bm{u}(\bm{x}_i(t), t),
 \label{eq:advection}
 }
 where $\bm{x}_i(t)$ is the position of the $i$-th particle at time $t$.
 We employ the 4-th order Runge-Kutta method for numerical integration,
 and the linear interpolation to obtain the velocity of each particles.
 Particle pairs are distributed homogeneously with relative separation
 of the grid scale $\Delta x$ at the initial time.

 \section{Results and discussion}

 \subsection{Fixed-time statistics}
 \begin{figure}
  \begin{center}
  \includegraphics[width=8.8cm]{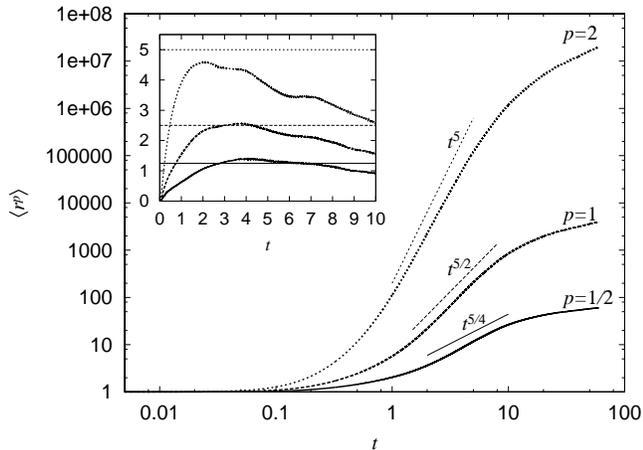}
  \end{center}
  \caption{\label{fig:ftime}
  Temporal evolution of mean relative separation $\langle r(t)^p\rangle$
  obtained in the case of NV11.
  Solid, dotted and dashed line refer to $p=1/2, 1,$ and $2$,
  respectively.
  Straight lines indicate Richardson's law.
  The inset shows the local slope of $\langle r(t)^p\rangle$.
  }
 \end{figure}

 \begin{figure*}[t] 
  \begin{center}
   \includegraphics[width=8.8cm]{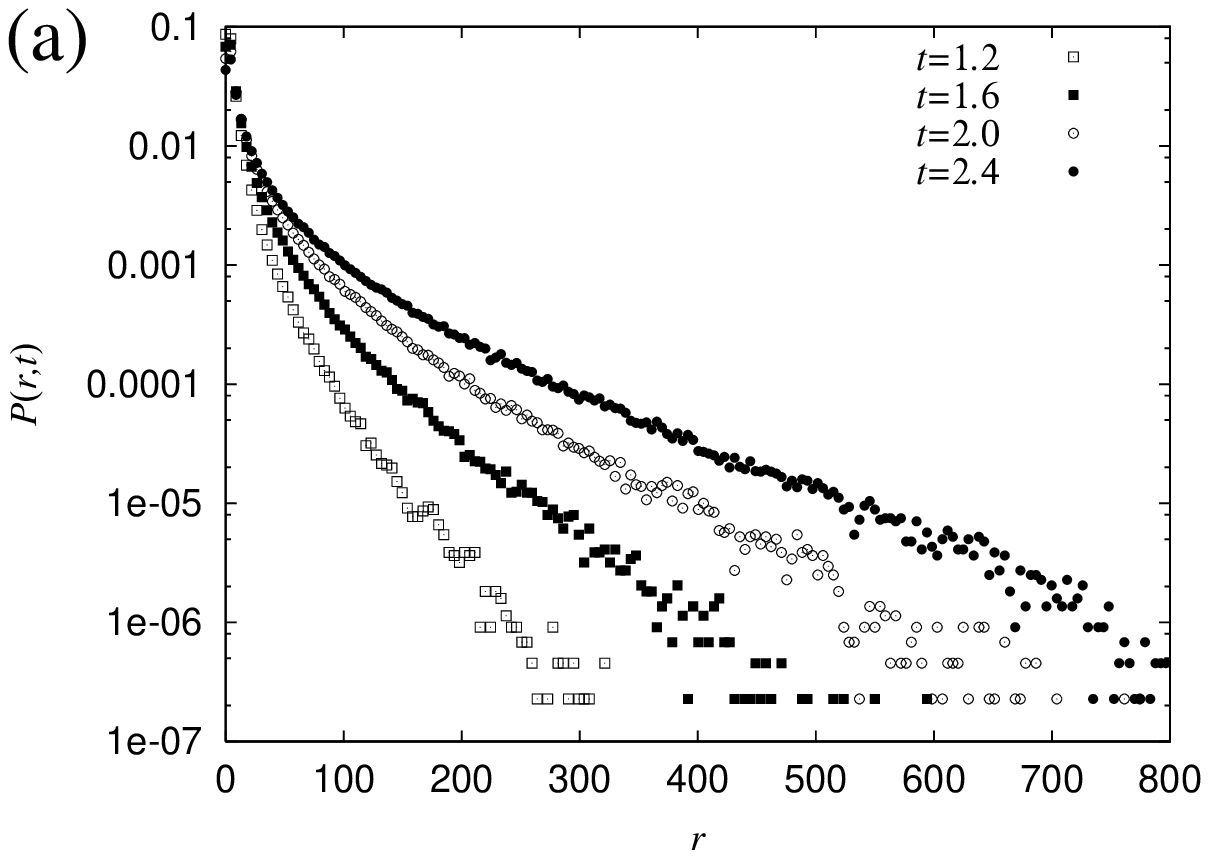}
   \includegraphics[width=8.8cm]{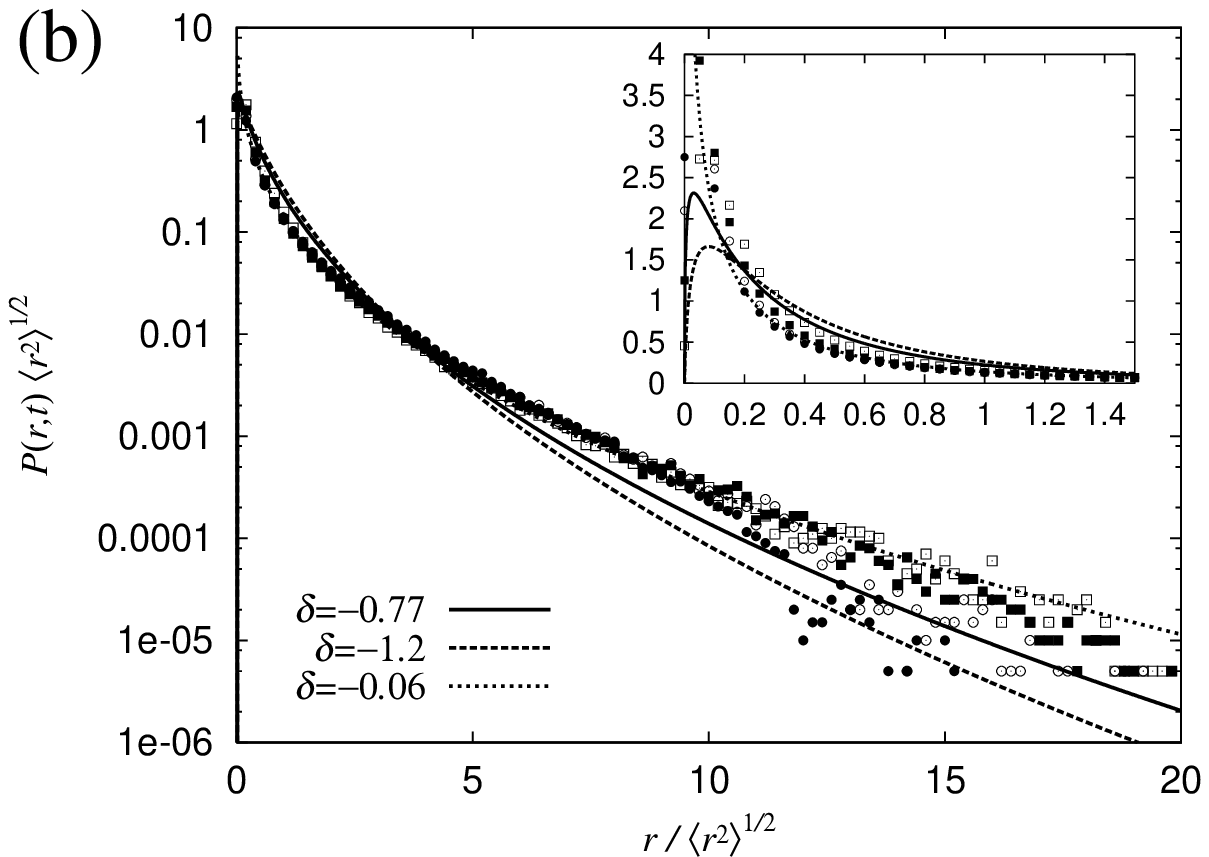}
  \end{center}
  \caption{\label{fig:fpdf}
  The PDF of relative separation $r$ obtained
  in the case of NV11.
  Different marks refer to different time:
  open box, closed box, open circle, and closed circle refer to $t=1.2$,
  $1.6$, $2.0$, and $2.4$, respectively.
  (a) the non-rescaled plot, (b) the rescaled plot with
  $\langle r^2\rangle^{1/2}$.
  The solid, dashed, and dotted lines are the similarity solutions of
  Palm's equation Eq.\ (\ref{eq:Palm-eq}) with
  $\delta=-0.77$,
  $-1.2$, and
  $-0.06$, respectively.
  The inset shows
  the linear plot of a blowup of the head region.
  }
 \end{figure*}

 First,
 we briefly
 discuss the
 results obtained by standard fixed-time statistics,
 which concerns the distribution of relative separation at a certain
 (fixed) time.

 In Fig.\ \ref{fig:ftime} we plot temporal 
 evolutions of
 the mean
 relative
 separation of particle pairs, $\langle r^p\rangle$,
 for different powers,
 $p=1/2$, $1$, and $2$.
 In the cases of $p=1/2$ and $1$,
 the generalized Richardson's 
 relations,
 $\langle r^p\rangle \propto t^{5p/2}$,
 are realized,
 though the times at which they start
 and the time intervals in which they hold differ.
 These differences result from
 the limited width of the inertial
 range in DNS.
 For example, in the case of NV11, the inertial range is roughly
 estimated as $80<r/\Delta x<320$.
 From
 Fig.\ \ref{fig:fpdf}(a), it is clear that the separation PDF
 broadens rapidly, so that
 $\langle r^p\rangle$ is
 contributed to by
 a quite broad range of
 relative separation
 including the dissipation, the inertial, and the energy containing
 ranges.
 As a result, the weight of each range varies with
 the power of the moment of the
 relative separation
 and thus,
 the range of time
 in which the
 contribution
 of the inertial range
 dominates differs accordingly.
 The slope of $\langle r^2\rangle$ is
 less steep than
 Richardson's law.
 This is because, in the case of $p=2$,
 the contribution of
 particle pairs
 in the energy containing range to
 $\langle r^p\rangle$
 is much larger  than that in the cases of $p=1$ and $1/2$. 
 In the energy containing range,
 relative separation process is described by the Brownian motion,
 $\langle r^2\rangle\propto t$,
 which is less steep than Richardson's law.
 Hence, the larger the contribution of the range is, the less
 steep the slope of $\langle r^p\rangle$ is.

 Even though the Richardson's law is observed
 for $p=1/2$ and $1$,
 the fact doesn't necessarily support
 the validity of the Richardson's law
 because it is reported that the temporal evolution of $\langle r^p\rangle$
 strongly depends
 on the initial separation of particle pairs \cite{BS2002a, GV2004}.
 Hence we have to check the scaling law by adopting a different method
 that is independent of the initial separation.

 In order to check self-similarity of the PDF of relative separation,
 we plot the PDF rescaled with $\langle r^2\rangle^{1/2}$
 in Fig.\ \ref{fig:fpdf}(b).
 The values of $\langle r^2\rangle^{1/2}$ at $t=1.2$, $1.6$, $2,0$, and
 $2.4$ are $15.3$, $28.6$, $47.5$, and $72.0\Delta x$, respectively.
 All PDFs collapse well for $r/\langle r^2\rangle^{1/2}<10$ and they are
 in good agreement with the similarity solution of Palm's equation with
 $\delta=-0.06$.
 However,
 we cannot conclude that the temporal evolution of the separation PDF is
 self-similar and governed by Palm's equation
 because Fig.\ \ref{fig:fpdf}(b) is obtained not from separations
 only in the inertial
 range but in
 the much wider range, $0<r<720\Delta x$.
 Although
 the collapse in Fig.\ \ref{fig:fpdf}(b) implies existence of
 a self-similar stage governed by Palm's equation,
 we have not had
 any reasonable explanation of the collapse.

 \subsection{Exit-time statistics}

   \begin{figure}
    \begin{center}
    \includegraphics[width=8.8cm]{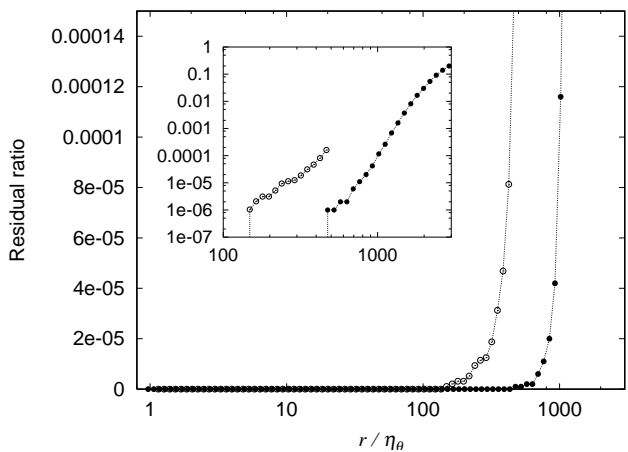}
    \end{center}
    \caption{\label{fig:erate}
    Residual ratio of particle pairs.
    Dotted lines with open and closed circles refer to
    data obtained by DNS at resolution
    $N=1024$ (NV10) and $N=2048$ (NV11) respectively.
    The inset is the log-log plot of the ratio.
    }
   \end{figure}

   \begin{figure}
    \begin{center}
    \includegraphics[width=8.8cm]{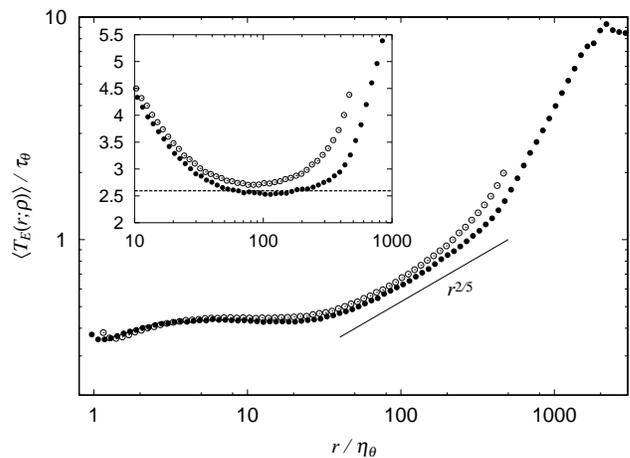}
    \end{center}
    \caption{\label{fig:etime}
    Scale dependence of 
    the mean exit-time $\langle T_E(\delta; \rho)\rangle$
    for $\rho=1.1$
    rescaled with the dissipation time scale, $\tau_\theta$.
    Open and closed circles refer to
    the results obtained by the present DNS at resolution
    $N=1024$ (NV10) and $N=2048$ (NV11), respectively.
    The inset is the compensated plot with
    $(\rho^\SE-1)(r/\eta_\theta)^{2/5}$.
    The dashed line represents estimated values of the coefficients
    $C_E^\mathrm{(BO)}\approx2.6$.
    }
   \end{figure}

   \begin{figure*}[t]
    \begin{center}
     \includegraphics[height=6.1cm]{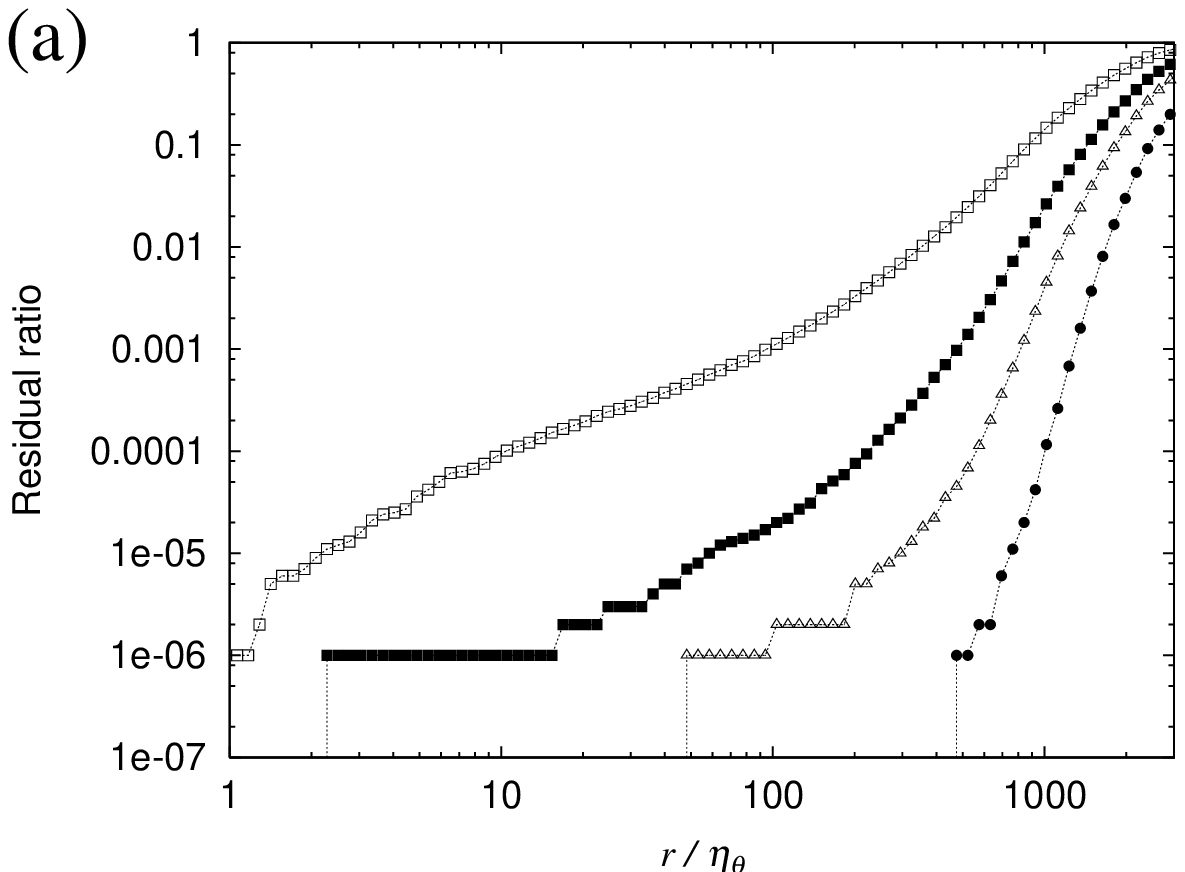}
     \includegraphics[height=6.1cm]{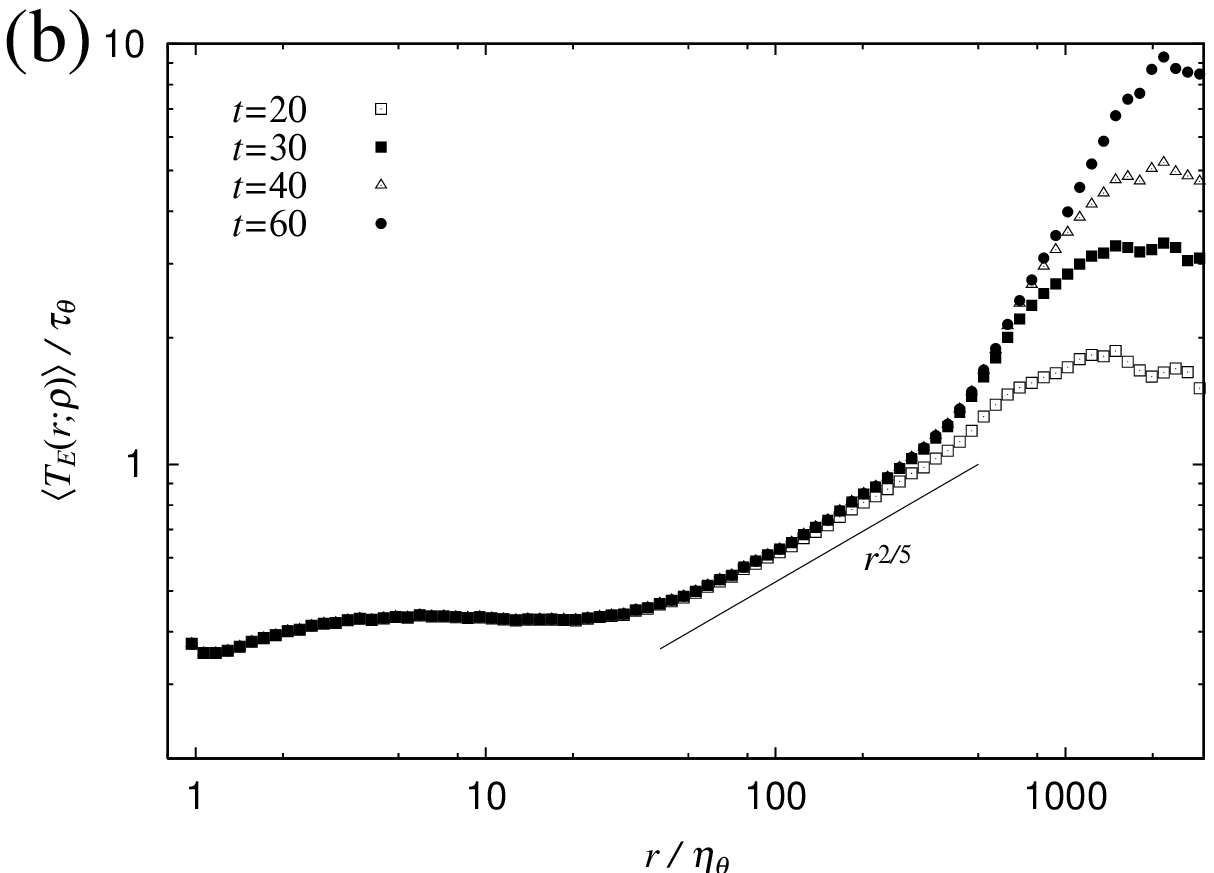}
    \end{center}
    \caption{\label{fig:marginal-etime}
    Mean exit-time and the residual ratio for several time steps
    ($\rho=1.1$):
    open box, closed box, open triangle, and closed circle represent
    $t=20$, $30$, $40$, and $60$, respectively.
    (a) the residual ratio for several time step,
    (b) the scale dependence of the mean exit-time.
    }
   \end{figure*}

   \begin{figure*}[t]
    \begin{center}
     \includegraphics[width=8.8cm]{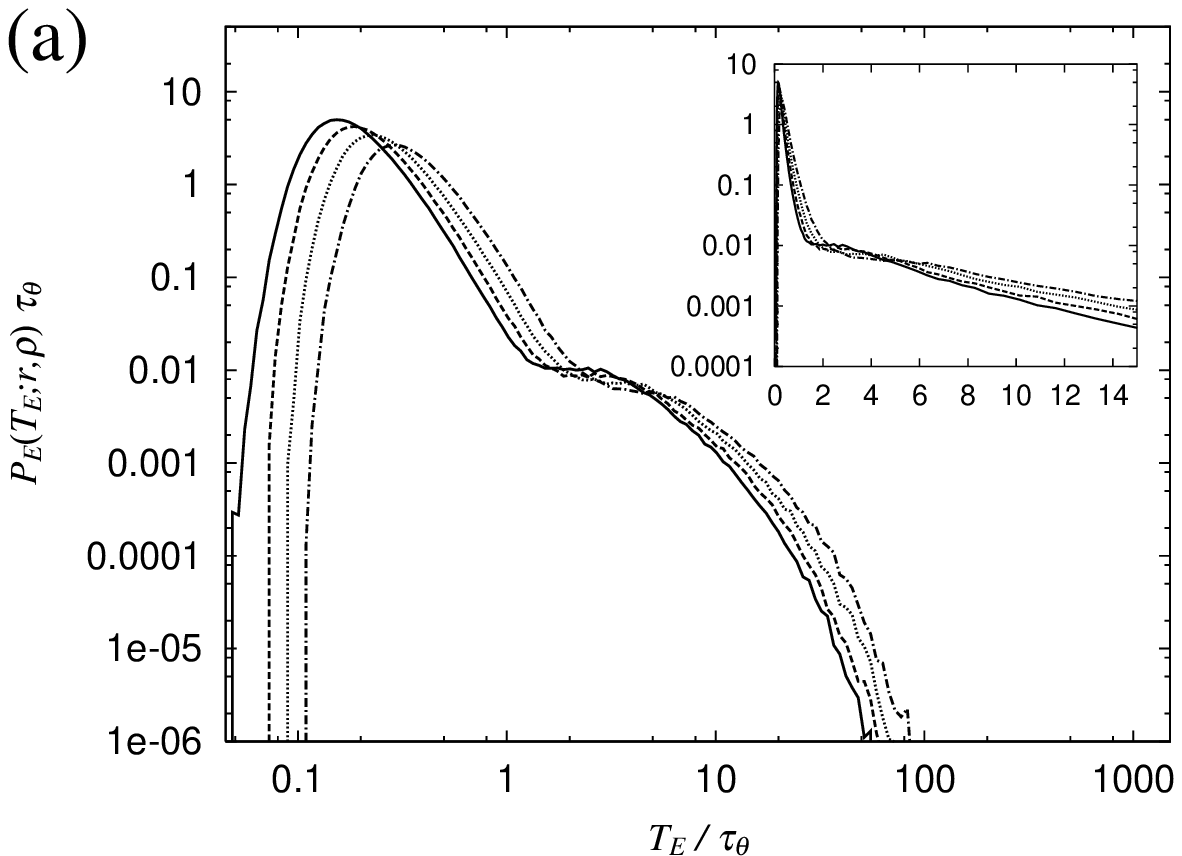}
     \includegraphics[width=8.8cm]{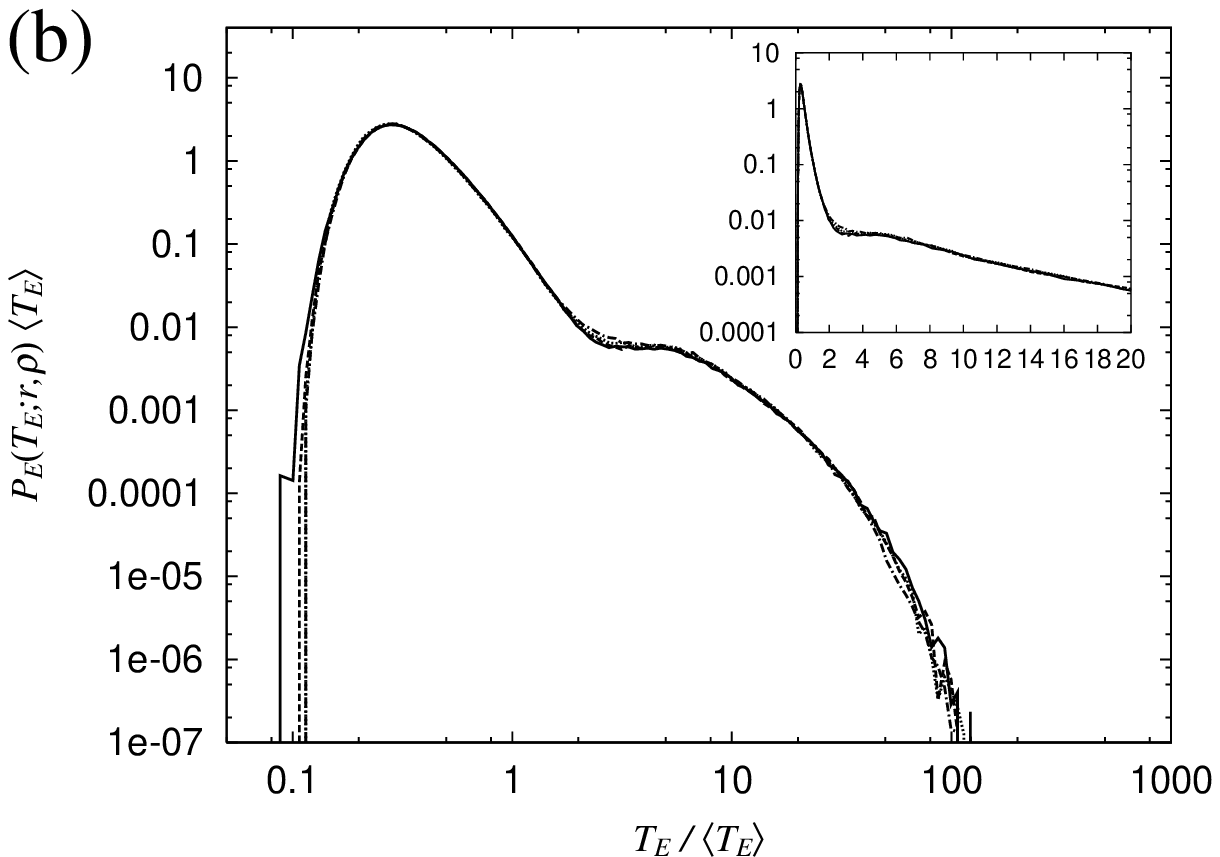}
    \end{center}
    \caption{\label{fig:pdf-et}
    The PDF of exit-time for $\rho=1.1$
    in the case of NV11.
    (a) the non-rescaled PDF, (b) the  rescaled PDF with
    the mean exit-time $\langle T_E(r;\rho)\rangle$.
    The solid, dashed, dotted and dashed-dotted
    lines refer to
    $r=80, 129, 208$ and, $335\Delta x$, respectively.
    The insets show a semi-log plot of the PDF.
    }
   \end{figure*}

   \subsubsection{Mean exit-time}
   The scale dependence of the mean exit-time obtained by our DNS
   is shown in Fig.\ \ref{fig:etime}.
   It is observed that, although the width is narrow, there is a region
   consistent with the Bolgiano-Obukhov scaling,
   $\langle T_E(r;\rho)\rangle\propto r^{2/5}$.
   The exit-time statistics is independent of initial separation of
   particle pairs if spatial scale $r$ is large enough
   for them to forget
   information of their initial conditions \cite{BS2002a,GV2004}.
   In our simulation, the initial separation of particle
   pairs is $\Delta x$ that
    is much smaller than the scales in which
    the scaling law holds.
   Therefore,
   this result indicates that
   Richardson's law is valid in the 2D-FC turbulence.

   According to the scaling law,
   the form of the mean exit-time is given by
   \EQA{
   \frac{\langle T_E(r;\rho) \rangle}{\tau_\theta}
   =
   C_E^\mathrm{(BO)}
   (\rho^\SE-1)
   \left(
   \frac{r}{\eta_\theta}
   \right)^\SE,
   \label{eq:etime-compensated}
   }
   where
   $C_E^\mathrm{(BO)}$
   is considered to be a universal constant.
   The inset of Fig.\ \ref{fig:etime} shows a compensated plot of the
   mean exit-time.
   Although
   the weak $\mathrm{Ra}_{\lambda}$ dependence of  $C_E^\mathrm{(BO)}$
   seen in Fig.\ \ref{fig:etime}
   requires higher resolution of  DNS to more accurate estimation,
   $C_E^\mathrm{(BO)}$ is estimated as $2.6$ in the present DNS.
   
   In order to obtain statistically reliable data of exit-time,
   it is important that the residual ratio of particle pairs must be small
   enough to take in slowly separating particle pairs.
   The residual ratio at a scale $r$ and a
time $t$ is defined as the
   ratio of
   particle pairs of which first passage times at $r$ are less than $t$.
   If the residual ratio is not small enough
   at the time when statistics
   of
   exit-time are calculated,
   the statistics cannot reflect slowly separating particle pairs.
   Figure \ref{fig:erate} shows the residual ratio of particle pairs used
   in the present work at the termination time of DNS, $t=60$.
   It is obvious that the
   ratio is almost zero in the inertial range,
   so that our exit-time data is reliable in the sense mentioned above.
   Figure \ref{fig:marginal-etime}(b) shows the scale dependence of the
   mean exit-time
   obtained from insufficient data and
   illustrates the   importance of the residual ratio.
   Although the results at $t=20$
   seem to have the wider inertial
   range than others, it is a fake.
   Therefore, to obtain statistically reliable data,
   we have to track particle pairs
   until
   the residual ratio becomes at least less
   than $0.1\%$.

   In contrast to the inertial range,
   the mean exit-time is almost constant in the
   dissipation range 
   ($r/\eta_{\vartheta} \lesssim 30$ in Fig.\ \ref{fig:etime})
   except in the initial transient 
   region, because the relative velocity is proportional to
   relative separation: $v(r)\propto r$. 
   The exit-time, then, is given by
   \EQA{
   \int_r^{\rho r}\frac{dr^\prime}{v(r^\prime)}
   \propto
   \int_r^{\rho r} \frac{dr^\prime}{r^\prime}
   =\log \rho=\mathrm{Const.}
   \label{eq:etime-constancy}
   }
   Note that the mean exit-times rescaled by $\tau_\theta$ 
   collapse each other  in Fig.\ \ref{fig:etime}.
   This is because there is only one characteristic time
   scale, $\tau_\theta$, in the dissipation range \cite{B1952b}.

   \subsubsection{PDF of exit-time and distribution of particle pairs in
   real space}
   
   We plot exit-time 
   PDF, $P_E(T_E; r, \rho)$,
   in the inertial range in Fig.\ \ref{fig:pdf-et}(a),
   and that rescaled by the mean 
   exit-time, $\langle T_E(r; \rho)\rangle$,
   in   Fig.\ \ref{fig:pdf-et}(b).
   Obviously the PDFs of different scales collapse onto one curve
   in Fig\ \ref{fig:pdf-et}(b).
   This means exit-time PDF is self-similar in the inertial range and
   indicates that relative dispersion process is self-similar.

\begin{figure*}
    \centerline{
    \includegraphics[height=9.8cm]{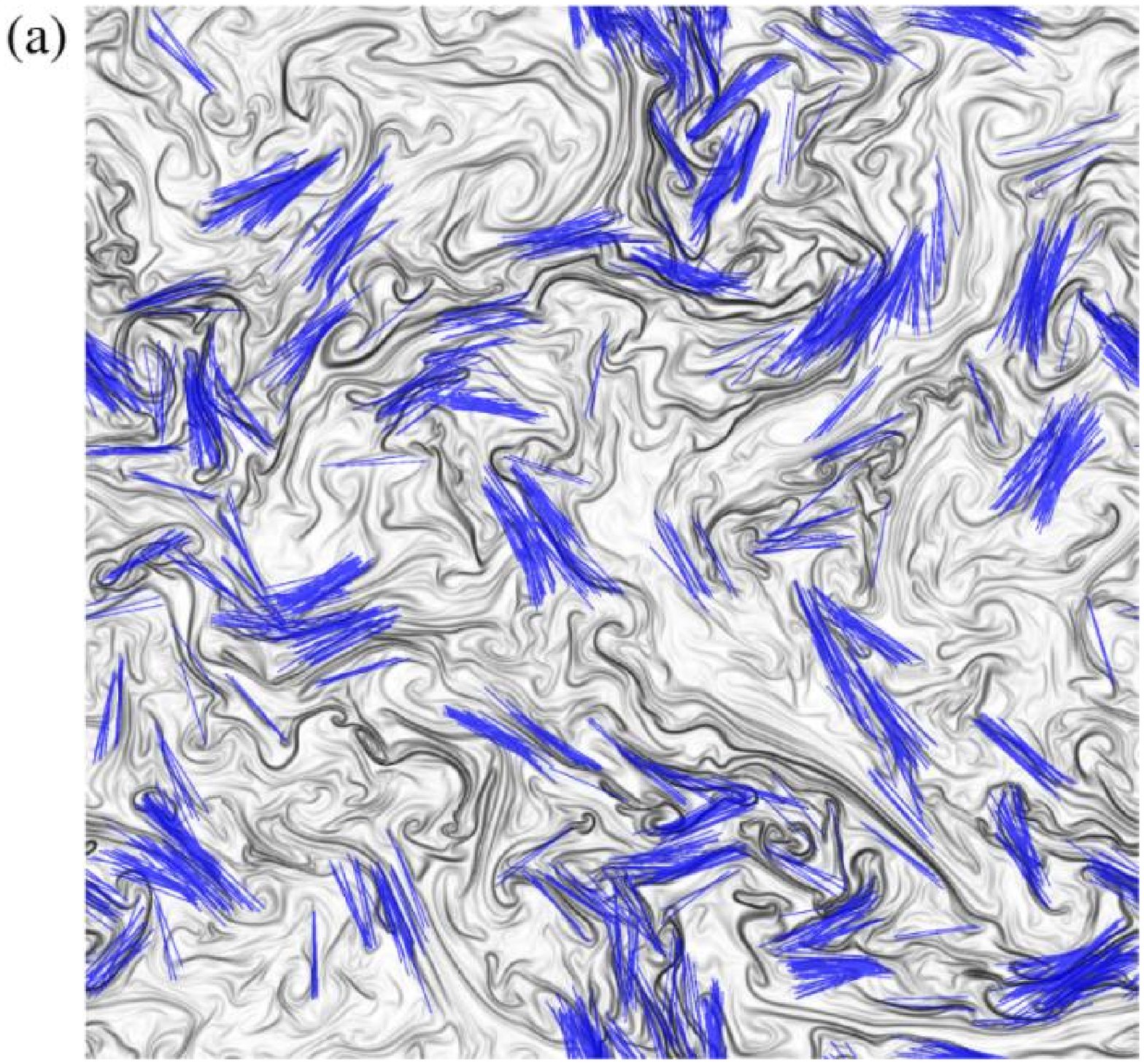}
    \hspace{.5mm}
    \includegraphics[width=7cm]{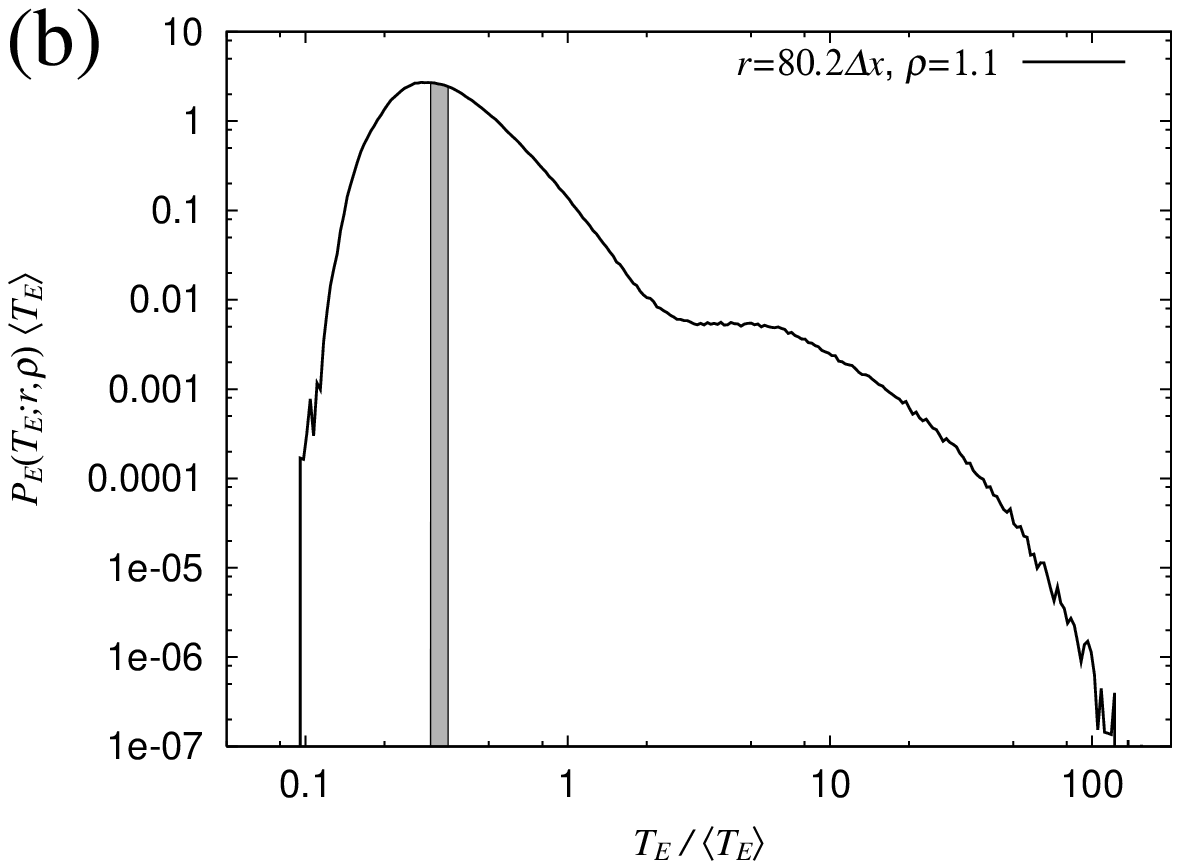}
    }
    \caption{\label{fig:pp-dist-I}
    Distribution of particle pairs in the Region-I,
    $0.3 < T_E/\langle T_E\rangle < 0.35$,
    for $r=80.2\Delta x$ and $\rho=1.1$ at $t=8$
    in the case of NV10.
    (a) the distribution of particle pairs superimposed onto a
    snapshot of the modulus
     of T-Vorticity, $|\bm{\chi}|$.
    Shading represents intensity of $|\bm{\chi}|$.
    A particle pair is represented by a line segment
    (particles are located at both ends of the line segment).
    The whole computational domain is shown.
    (b) the PDF of exit-time rescaled with
    the mean.
    The particle pairs drawn in (a) belong to the  gray 
    region,  $0.3 < T_E/\langle T_E\rangle < 0.35$.
    }
   \end{figure*}

\begin{figure*}
    \centerline{
    \includegraphics[height=9.8cm]{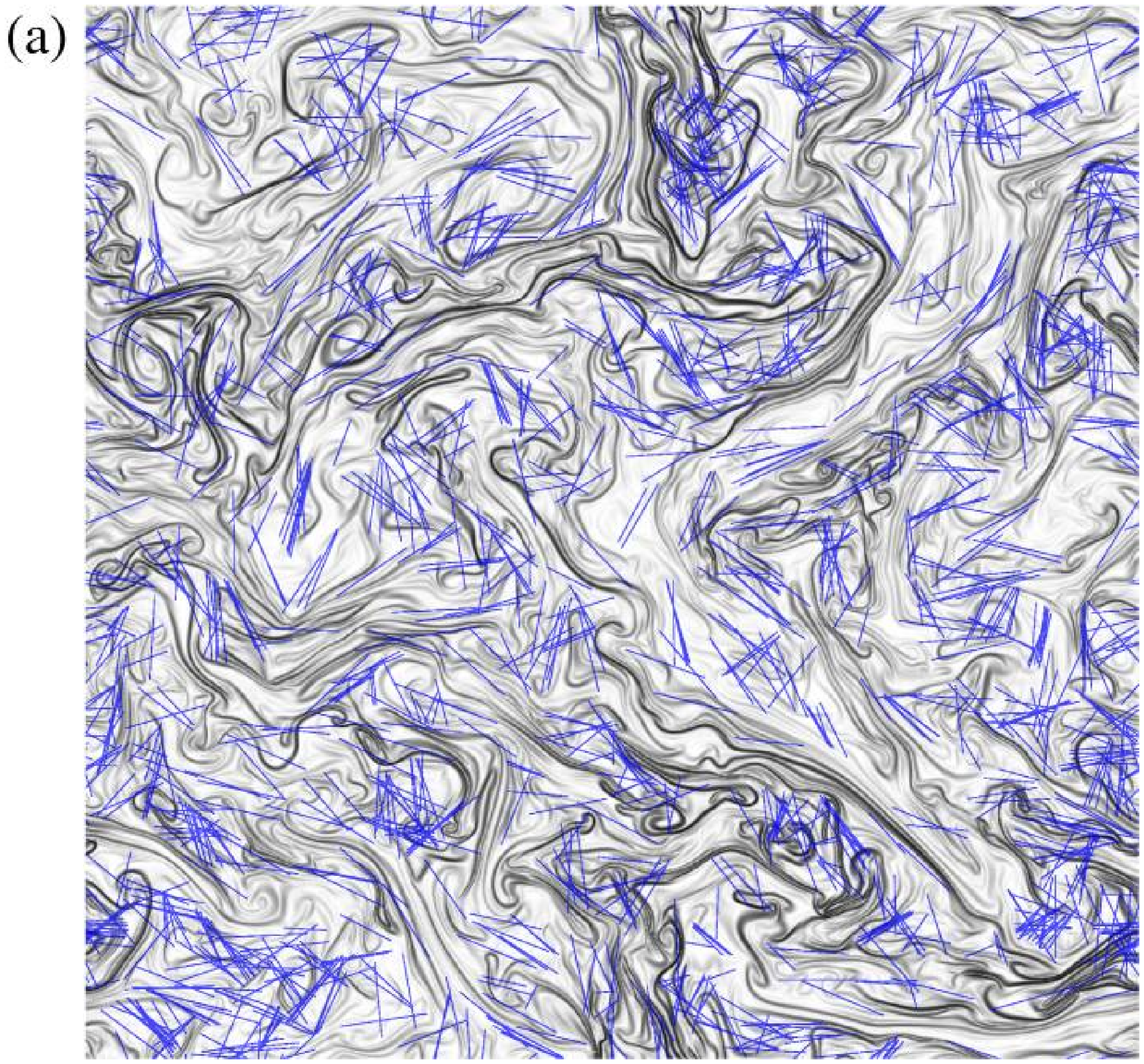}
    \hspace{.5mm}
    \includegraphics[width=7cm]{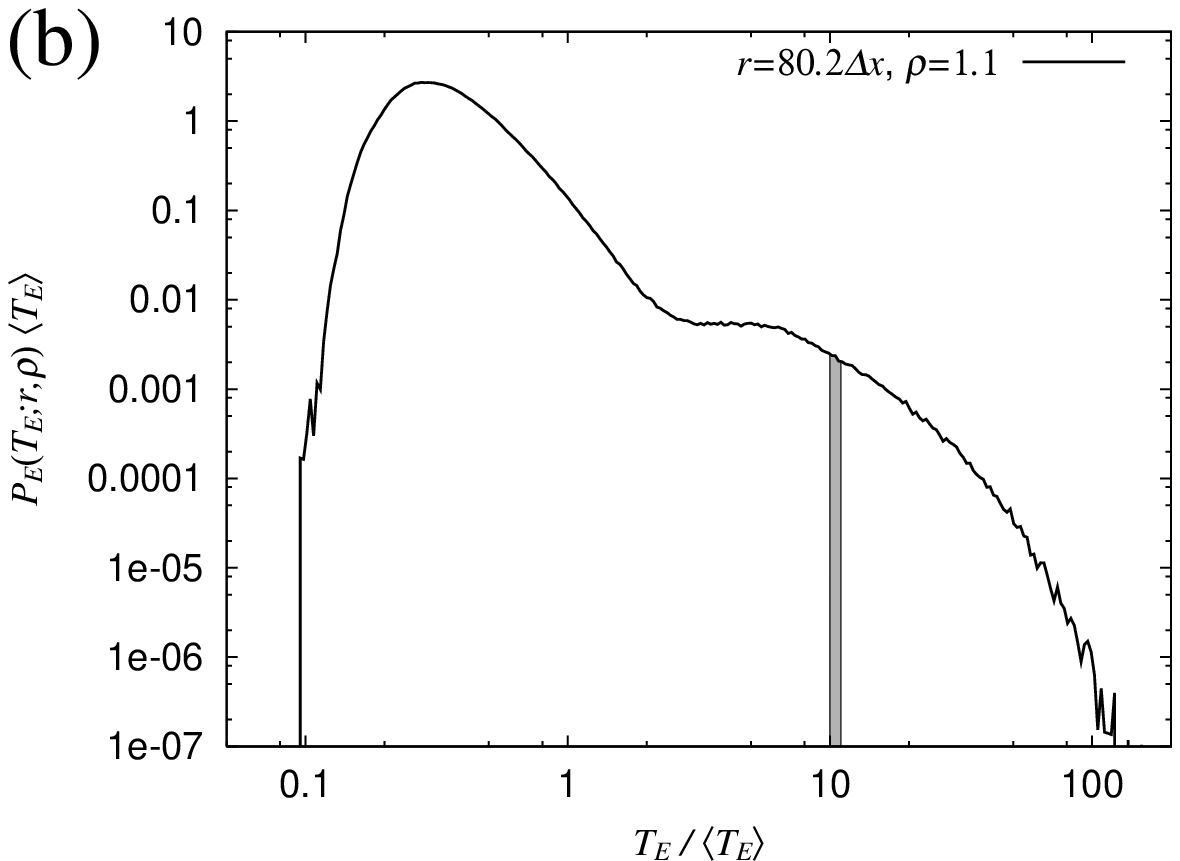}
    }
    \caption{\label{fig:pp-dist-II}
    Distribution of particle pairs in the Region-II,
    $10 < T_E/\langle T_E\rangle < 11$, for $r=80.2\Delta x$ and
    $\rho=1.1$ at $t=8$ in the case of NV10.
    (a) and (b) are drawn in the same manner as 
    Fig.\ \ref{fig:pp-dist-I}.
    }
   \end{figure*}

   As is clear from Fig.\ \ref{fig:pdf-et}(b), the exit-time PDF
   consists of two regions: the sharp peak and the long exponential tail.
   We call these two regions the Region-I and
   the Region-II, respectively.
   The qualitative difference
   in
   form between these indicates that
   the PDF reflects two different types of motions.
   In order to clarify the difference,
   snapshots of typical distribution of particle pairs
   in the Region-I and -II are shown
   in Figs.\ \ref{fig:pp-dist-I} and \ref{fig:pp-dist-II},
   respectively,
   with a snapshot of
   the magnitude of T-Vorticity field $|\bm{\chi}|$.
   
   \begin{figure}
    \begin{center}
    \includegraphics[width=8.8cm]{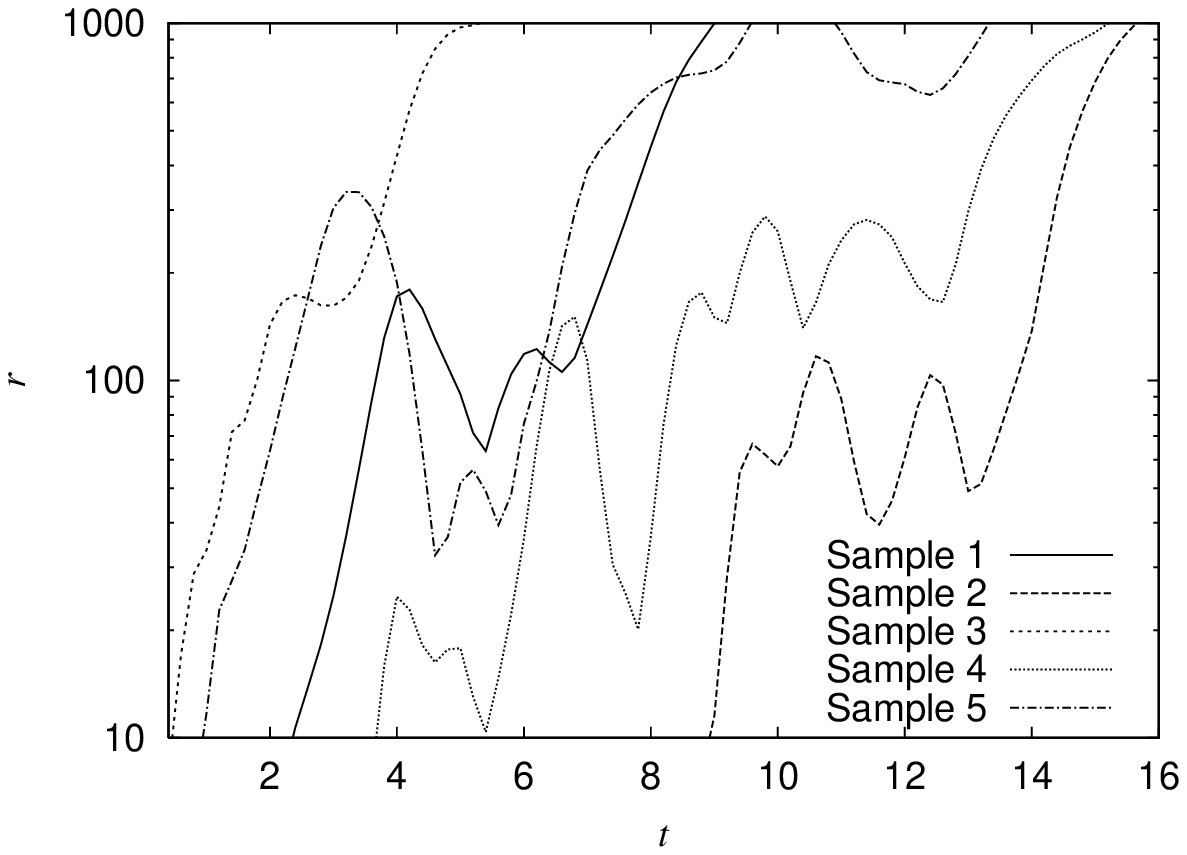}
    \end{center}
    \caption{\label{fig:PP}
    Temporal evolution of relative separation for several particles
    in the case of NV11.
    Different lines represent different particle pairs.
    }
   \end{figure}

   In Fig.\ \ref{fig:pp-dist-I}(a),
   on the snapshot of $|\bm{\chi}|$ at $t=8$ ($\equiv
   t_s$),
   we plot line segments 
   representing particle pairs 
   of which
   exit-time $T_E(r_s;\rho_s)$
   for $r_s= 80.2\Delta x$ and $\rho_s=1.1$
   satisfies   $0.3<T_E/\langle T_E(r_s; \rho_s)\rangle<0.35$, 
   that is, $T_E(r_s;\rho_s)$ is in the shaded region 
   in Fig. \ref{fig:pp-dist-I}(b).
   In order to select such particle pairs,
   we first extract pairs satisfying the condition
   $0.3<T_E/\langle T_E(r_s; \rho)\rangle<0.35$,
   and then pick out ones of which first passage times at $r_s$ are
   smaller
   than $t_s$ and those at $\rho r_s$ are
   larger
   than $t_s$.
   Figure \ref{fig:pp-dist-II}(a) is drawn by the same procedure.
   To draw Figs.\ \ref{fig:pp-dist-I} and \ref{fig:pp-dist-II},
   we used the data of NV10.

   In 2D-FC turbulence, fine coherent structures are well approximated by
   the Burgers T-Vortex layer, so that shear layers are formed around 
    the  structures \cite{TM2003}.
   In addition, such structures are persistent in time to some extent.
   Hence,
   particles around the coherent structures are advected along them
   and
   relative separations of the particles expand or compress persistently.
   We call this type of motion the persistent separation.
   As is shown in Fig.\ \ref{fig:pp-dist-I}(a),
   particle pairs in the Region-I
   appear to
   be along  the fine coherent structures.
   Moreover relative separations of particle pairs
   in Fig.\ \ref{fig:pp-dist-I}(a)
   expand rapidly as can be known by their short exit-time.
   Therefore,
   Fig.\ \ref{fig:pp-dist-I} supports
   the picture of persistent separations.

   In contrast to the Region-I, the distribution of particle pairs in the
   Region-II is
   in disorder
   (see Fig.\ \ref{fig:pp-dist-II}(a)).
   Because the particle pairs
   contained in
   the Region-II have long exit-time,
   the longitudinal relative velocity of them remains small positive
   during the passage from
   $r_s$ to $\rho r_s$ 
   or
   moving direction of relative separation of the pairs changes from
   expansion to compression
   at least once before they
   reach $\rho r_s$.
   However it is quite unlikely that relative velocity remains small
   positive for a long time, so that the latter is
   probably
   the main mechanism of
   the formation of
   the Region-II.
   Hence, it is expected that  
   both of expanding and compressing pairs are contained
   in Fig.\ \ref{fig:pp-dist-II}(a).
   In fact, there are several pairs placed along coherent structures,
   which 
   probably are
   expanding.
   Note that their relative separations
   are not confined to be larger than $r_s$.
   In addition, there are also pairs
   across the structures,
   which are
   presumably compressing
   because the structures are approximated by the
   Burgers T-Vortex layer.
   Moreover, some pairs are located
   at positions where the fine coherent structures twist.
   The twisted
   regions
   are generated
   when well-stretched structures lose their activities and are folded,
   or when the structures are
   generated by plumes.
   Therefore
   the particle pairs in such regions are
   changing their moving directions from expansion to compression or the
   opposite.

   Figure \ref{fig:PP} shows five typical evolutions of
   particle-pair separation. Sample 3 expands persistently without any
   strong compression but Sample 4 experiences both persistent 
   expansion and  compression several times. The former corresponds to
   motions along structures and belongs to the Region-I; 
   the latter does across  structures or twisted region and belongs 
   to the Region-II. Note that even a single evolution contains 
   motions that are categorized  into the Region-I or into 
   the Region-II.
   These
   facts
   are consistent with 
   the assumptions
   of the self-similar
   telegraph model: relative separation process consists of persistent
   expansion and compression with some random transition mechanisms.

   \subsubsection{Estimation of $\delta$}
   
   \begin{figure}
    \begin{center}
    \includegraphics[width=8.8cm]{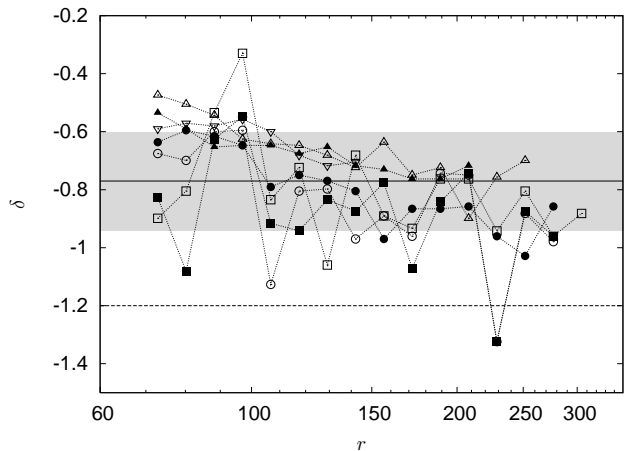}
    \end{center}
    \caption{\label{fig:delta}
    Estimated values of $\delta$ in the case of NV11.
    Dotted lines with open box, closed box, open circle, closed circle,
    open triangle, closed triangle, and reverse open triangle
    denote
    $\rho=1.1^{1/8}$, $1.1^{1/4}$, $1.1^{1/2}$, $1.1$, $1.1^2$, $1.1^4$,
    and $1.1^8$, respectively.
    Solid and dashed lines indicate the mean value, $\delta=-0.77$,
    and the value of the Richardson's case, $\delta=-1.2$, respectively.
    The shaded region represents the standard deviation.
    }
   \end{figure}

   It is expected that
   expansion and compression  of relative separation differ 
   in persistency; the former is experienced mainly by particle pairs along 
   coherent structures and the latter does by those across the structures.
   Since  the auto-correlations of strain and temperature along
   the structures have  longer characteristic lengths  than those 
   across them in 2D-FC turbulence \cite{TM2003},
   expanding motions must be more persistent than compressing ones.
   So as to confirm this consideration, we investigate the slope of
   the tail of the exit-time PDF. Then,
   with Eq.\ (\ref{eq:Palm-etime}),
   we estimate  $\delta$
   that
   describes the difference in persistency
   between expansion and compression of relative separation.

   Note that
   using the asymptotic form of the exit-time PDF
   calculated from Palm's equation, Eq.\ (\ref{eq:Palm-etime}),
   may be justified as follows.
   Introducing a time scale $\bar{t}$,
   the l.h.s.\ of the self-similar telegraph model, Eq.\ (\ref{eq:T-Model}),
   is rewritten as follows:
   \EQA{
   \mathrm{l.h.s.}
   =
   \frac{T_c(r)}{\lambda \bar{t}}\frac{1}{\bar{t}}
   \frac{\partial^2 P}{\partial \hat{t}^2}
   +
   \frac{1}{\bar{t}}
   \frac{\partial P}{\partial \hat{t}},
   }
   where $\hat{t}=t/\bar{t}$.
   In the case that $\bar{t}\gg T_c(r)/\lambda$ 
   at a certain spatial scale $r$,
   the first term of the l.h.s.  can be neglected, 
   and thus, the model is reduced to Palm's equation.
   Hence, the tail of the exit-time PDF by the telegraph model,
   which consists of slowly-separating particle pairs,
   must
   coincide
   with that by Palm's equation.
   It is also easy to show that
   the mean exit-time is the same between
   Eqs.\ (\ref{eq:T-Model}) and (\ref{eq:Palm-eq}).
   We, therefore, adopt
   Eq.\ (\ref{eq:Palm-etime}) for
   the
   estimation of $\delta$.
   
   Figure \ref{fig:delta} shows the estimated values of $\delta$ in the
   inertial range.
   The mean value of $\delta$ is $-0.77$
   and the standard deviation is $\pm0.17$.
   This negative value denotes that expanding motions are more 
   persistent than compressing ones,
   which supports
   the above expectation.

   In addition,
   $\delta$ is larger than
   that of the Richardson case (the zero-drift),
   $\delta=-1.2\equiv\delta_0$.
   Because $\delta=\lambda^+ - \lambda^-$,
   $\delta>\delta_0$ means
   $\lambda^- < \lambda^+ - \delta_0 \equiv \lambda^-_0$,
   that is, the compressing motion of relative separation
   is more persistent than that of the Richardson case.
   This fact
   results in the negative drift in the self-similar telegraph
   model, Eq.\ (\ref{eq:T-Model}).
   The negativity of the drift
   can be accepted considering
   that coherent structures in
   2D-FC turbulence are string-like
   and scale-transversal ones \cite{TM2003}.
   That is, because the coherent structures
   in 2D-FC turbulence are not nested in scales,
   there are few obstacles to compressing motions.
   On the other hand,
   the typical structures
   in 2D-IC turbulence are nested
   vortices called ``cat's eye in a cat's eye''.
   A separation process of particle pairs
   in 
   2D-IC turbulence, thus, is a
   step-by-step one 
   each step of which consists of
   a trapping by one of nested vortices and  sudden separation 
   into  a next larger vortex \cite{GV2004};
   compressing motions are probably blocked by the nested vortices.
   In fact,
   the estimated values of $\delta$ from DNS of 2D-IC turbulence
   by Boffetta \& Sokolov \cite{BS2002a} and Goto \& Vassilicos
   \cite{GV2004} are smaller than that of the Richardson case 
   \cite{OT2006b};
   the compressing motion of relative separation is
   less persistent than the Richardson case.
   This results in the positive drift in Eq.\ (\ref{eq:T-Model}).
   We, therefore,  conclude
   that the drift term in the
   self-similar telegraph model, Eq.\ (\ref{eq:T-Model}), reflects the
   characteristics of coherent structures of the flow.

   \subsubsection{Comparison of the Region-II with a solution of Palm's equation}
   
   \begin{figure}
    \begin{center}
    \includegraphics[width=8.8cm]{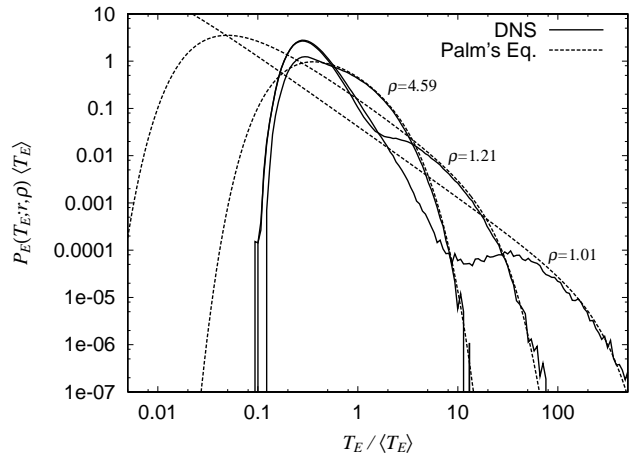}
    \end{center}
    \caption{\label{fig:epdf-palm}
    Rescaled PDF of exit-time ($r=107\Delta x$)
    in the case of NV11 and the prediction of Palm's
    equation ($\delta=-0.77$) for several values of $\rho$.
    Solid and dashed lines refer to the results of DNS and
    that obtained from Palm's
    equation through the relation
    Eq.\ (\ref{eq:palm-epdf}).
    }
   \end{figure}

   In Fig.\ \ref{fig:epdf-palm}, we compare the PDF of exit-time
   obtained by DNS with that calculated from Palm's equation by
   Eq.\ (\ref{eq:palm-epdf}).
   In the Region-I, the form of the PDF by DNS is totally different from that
   by Palm's equation.
   In the Region-II, however,  the two PDFs collapse onto a single curve
   even when
   $\rho$ is very small,
   and thus, the
   separation process of particle pairs in the Region-II can be
   described by Palm's equation, Eq.\ (\ref{eq:Palm-eq}).
   This indicates that the motion of particle pair is diffusive
   in the Region-II;
   the separating process from $r\rho^{-1}$ to $r$ doesn't affect
   that from $r$ to $r\rho$.
   It is also observed that the form of the exit-time PDF varies
   depending on the value of $\rho$ in Fig.\ \ref{fig:epdf-palm}.
   The larger the value of $\rho$ is, the larger the
   proportion of the
   Region-II becomes.
   If $\rho$ is sufficiently large,
   the whole PDF seems to be occupied by
   the Region-II and the relative separation process is substantially
   described by Palm's equation, Eq.\ (\ref{eq:Palm-eq}).

   As mentioned in the previous
   subsection,
   the tail of the exit-time PDF calculated from the self-similar telegraph
   model, $P_E^\mathrm{(T)}(T_E;r,\rho)$, 
   must agree with
   that
   calculated from Palm's equation, $P_E^\mathrm{(P)}(T_E;r,\rho)$.
   In addition,
   as the value of $\rho$ gets larger,
   the region 
   of $P_E^\mathrm{(T)}(T_E;r,\rho)$
   overlapping
   with $P_E^\mathrm{(P)}(T_E;r,\rho)$ 
   expands.
   This is because,
   for large values of $\rho$,  most of exit-times of particle pairs  
   are longer than the characteristic time $T_c(r)/\lambda$.
   Hence, the self-similar telegraph model 
   probably has
   the same properties  as the 
   results shown in Fig.\ \ref{fig:epdf-palm}.

   \subsubsection{Characteristics of the Region-I and estimation of $\lambda$}
   \begin{figure}
    \begin{center}
    \includegraphics[width=8.8cm]{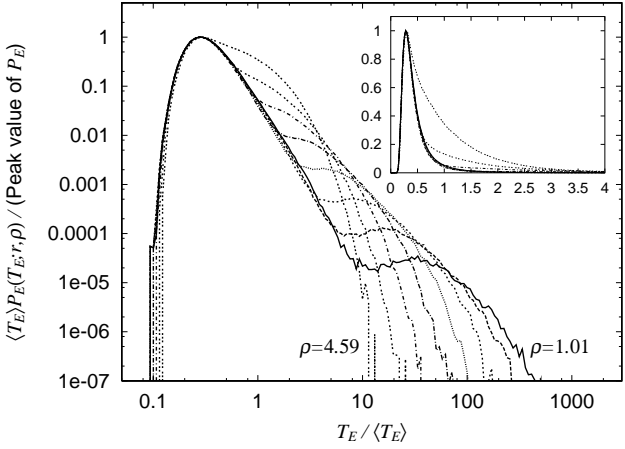}
    \end{center}
    \caption{\label{fig:epdf-eqpeak}
    Rescaled PDF of exit-time ($r=107\Delta x$)
    in the case of NV11.
    The peak value is normalized to be unity.
    Different lines represent different values of $\rho$:
    $\rho=4.59$, $2.14$, $1.46$, $1.21$, $1.1$, $1.05$, $1.02$, and
    $1.01$ from the curve with the lowest cut-off rescaled exit-time to
    that with the highest.
    The inset is the same plot in linear scale.
    }
   \end{figure}

   Figure \ref{fig:epdf-eqpeak} shows the exit-time PDF
   for NV11
   rescaled with $\langle T_E(r;\rho)\rangle$ and normalized by
   their peak values.
   It is clear that the shape of the Region-I is independent of $\rho$.
   That is, there exists a $\rho$-independent
   function representing the shape of the
   exit-time PDF in the Region-I,
   $\widehat{P}_E^\mathrm{(I)}(\widetilde{T}_E)$.
   Here,
   $\widetilde{\ \ }$ denotes rescaling by the mean exit-time,
   $\langle T_E(r;\rho)\rangle$.
   We assume that $\widehat{P}_E$ is normalized as
   $\int_0^\infty \widehat{P}_E(\widetilde{T}_E) d\widetilde{T}_E=1$.
   Then, in the Region-I, the exit-time PDF is written as
   \EQA{
   \widetilde{P}_E(\widetilde{T}_E;\rho) =
   W(\rho) \widehat{P}_E^\mathrm{(I)}(\widetilde{T}_E),
   }
   where $W(\rho)$ is a normalization factor depending on $\rho$.
   The exit-time PDF, $\widetilde{P}_E(\widetilde{T}_E;\rho)$,
   takes the  peak value at $\widetilde{T}_E^\mathrm{max}$.
   It is estimated that $\widetilde{T}_E^\mathrm{max}\approx0.3$
   in Fig.\ \ref{fig:epdf-eqpeak},
   which
   can be regarded as
   a characteristic time scale 
   of the Region-I.
   As shown in the inset of Fig.\ \ref{fig:epdf-eqpeak},
   $\widehat{P}_E^\mathrm{(I)}(\widetilde{T}_E)$ distributes sharply around
   $\widetilde{T}_E^\mathrm{max}$.

   If the Region-I is formed by particle pairs of which relative
   separations expand persistently according to $v(r)=\CA r^{1-\SE}$,
   the exit-time in the Region-I, $T_E^\mathrm{(I)}(r;\rho)$, is given by the
   pass through time
   from $r$ to $\rho r$:
   \EQA{
   T_E^\mathrm{(I)}(r;\rho)=\int_r^{\rho r}\frac{dr}{v(r)}
   =\frac{(\rho^\SE-1)r^\SE}{\SE\CA}.
   }
   Then,
   using the general form of the mean exit-time,
   Eq.\ (\ref{eq:etime-general}),
   the exit-time rescaled by $\langle T_E(r;\rho)\rangle$ 
   in the Region-I,
   $\widetilde{T}_E^\mathrm{(I)}(r;\rho)$, is
   \EQA{
   \widetilde{T}_E^\mathrm{(I)}(r;\rho)\equiv
   \frac{T_E^\mathrm{(I)}(r;\rho)}{\langle T_E(r;\rho)\rangle}
   =
   \frac{1}{\SE \check{C}_E \CA},
   \label{eq:T-Model_Region-I}
   }
   which is independent of $\rho$.
   Thus, $\widehat{P}_E^\mathrm{(I)}(\widetilde{T}_E)$ can be connected
   to the PDF of $\CA$,
   $P_{\CA}(\CA)$, which is the PDF of
   Lagrangian relative
   velocity, $v(r)$, rescaled with $r^{1-\SE}$:
   \EQA{
   \widehat{P}_E^\mathrm{(I)}(\widetilde{T}_E)=
   \widehat{P}^\mathrm{(I)}_E
   \left(
   \frac{1}{\SE \check{C}_E \CA}
   \right)\equiv
   \SE \check{C}_E
   \CA^2 P_{\CA}(\CA).
   }
   Note that
   $P_{\CA}(\CA)$
   is expected to be independent both of $r$ and
   $\rho$.
   Since, in the self-similar telegraph model, 
   the distribution of the coefficient, $\CA$, of the relative  velocity
   is not considered,  $P_{\CA}(\CA)$  is a $\delta$-function:
    $P_{\CA}(\CA)=\delta(\CA-\CA_c)$.
   Accordingly, the Region-I of the exit-time PDF is
   approximated by a
   $\delta$-function at 
   $\widetilde{T}_E^\mathrm{(I)}$ in the model.

   Combining Eqs.\ (\ref{eq:avetime-palm}) and (\ref{eq:T-Model_Region-I}),
   we can estimate $\lambda$ from $\widetilde{T}_E^\mathrm{(I)}$.
   From Eq.\ (\ref{eq:avetime-palm}),
   $\check{C}_E$ in the case of the self-similar telegraph model is given by
   \EQA{
   \check{C}_E = \frac{\lambda}{\CA}
   \frac{1}{\SE(2\SE-\delta)}.
   }
   Then, $\lambda$ is calculated as follows:
   \EQA{
   \lambda =
   \frac{2\SE-\delta}{\widetilde{T}_E^\mathrm{(I)}}.
   }
   If we assume that
   $\widetilde{T}_E^\mathrm{(I)}\approx\widetilde{T}_E^\mathrm{max}
   \approx0.3$,
   then,
   $\lambda$ is estimated as $5.2$.
  However, the value of $\widetilde{T}_E^\mathrm{(I)}$ is  
   not necessarily well defined. Roughly speaking, it may take a value satisfying   
   $0.19 \lesssim \widetilde{T}_E^\mathrm{(I)}
   \lesssim 0.45$,
   if assuming $\widetilde{T}_E^\mathrm{(I)}$ is 
   in the half-value width of the Region-I around 
   $\widetilde{T}_E^\mathrm{max}$. 
   Then,  the estimated value of $\lambda$  is  in the range, 
   $3.5 \lesssim \lambda \lesssim 8.3$.

 \section{concluding remarks}

  We have investigated relative dispersion
  in 2D free convection turbulence by direct numerical simulation.
 In the inertial range, where the entropy cascade dominates,
 we have confirmed 
 with exit-time statistics that relative 
 dispersion  satisfies the Bolgiano-Obukhov scaling and,   
 therefore, is self-similar. 
 It was also shown that the exit-time PDF, $P_E(T_E;r,\rho)$,
 is divided into two parts, the Region-I and -II,
 and that both of them satisfy the scaling-law and the self-similarity.
 $P_E(T_E;r,\rho)$
 is written as the following form:
 \EQA{
 P_E(T_E;r,\rho)
 \EQAEQ
 H(T_E^{\mathrm{div}}(r;\rho)-T_E) P_E^\mathrm{(I)}(T_E;r,\rho)
 \qquad\quad
 \nonumber
 \\
 \EQAMK + \EQAMK
 H(T_E-T_E^{\mathrm{div}}(r;\rho)) P_E^\mathrm{(II)}(T_E;r,\rho),
 }
 where
 $T_E^{\mathrm{div}}(r;\rho)$ is a division time-scale (exit-time)
 between the Region-I and
 -II, and
 $H(x)$ is a smoothed step function such that $H(x)=0$ if $x\ll0$ and
 $H(x)=1$ if $x\gg0$.
 $P_E^\mathrm{(I)}(T_E;r,\rho)$ and $P_E^\mathrm{(II)}(T_E;r,\rho)$ correspond to
 the exit-time PDF of the Region-I and -II, respectively.
 The investigation of the distribution of particle pairs 
 in the real space indicates that
 the Region-I and -II are formed, respectively, by
 particle pairs expanding along coherent structures
 and by those experiencing turns between
 expansion  and compression
 (Figs.\ \ref{fig:pp-dist-I} and \ref{fig:pp-dist-II}).

 Figures \ref{fig:epdf-eqpeak} and \ref{fig:epdf-palm}
 show the following characteristics of $P_E^\mathrm{(I)}(T_E;r,\rho)$
 and $P_E^\mathrm{(II)}(T_E;r,\rho)$.
 The form of $P_E^\mathrm{(I)}(T_E;r,\rho)$ is independent of $\rho$ if
 it is rescaled with the mean exit-time.
 Moreover, if we assume that the Region-I is
 constituted
 by
 persistently-separating particle pairs,
 the form of $P_E^\mathrm{(I)}(T_E;r,\rho)$ is related to the PDF of
 Lagrangian relative velocity, $v(r)=\CA r^{1-\SE}$.
 That is,
 \EQA{
 \widetilde{P}_E^\mathrm{(I)}(\widetilde{T}_E;\rho)
 =
 \frac{W(\rho)}{\SE \check{C}_E \widetilde{T}_E^2}
 P_{\CA}\left(
 \frac{1}{\SE \check{C}_E \widetilde{T}_E}
 \right),
 }
 where $W(\rho)$ is a $\rho$-dependent normalization factor,
 and $P_{\CA}(\CA)$
 is the PDF of $v(r)$ rescaled with $r^{1-\SE}$.
 On the other hand,
 $P_E^\mathrm{(II)}(T_E;r,\rho)$ agrees with the exit-time PDF calculated
 from Palm's equation, $P_E^\mathrm{(P)}(T_E;r,\rho)$,
 the form of which varies depending on $\rho$ even when it is rescaled
 with $\langle T_E(r;\rho)\rangle$.
 Hence
 $\widetilde{P}_E(\widetilde{T}_E;\rho)$
 is written as
 \EQA{
 \widetilde{P}_E(\widetilde{T}_E;\rho)
 \EQAEQ
 \widetilde{H}(\widetilde{T}_E^{\mathrm{div}}(\rho)-\widetilde{T}_E)
 \frac{W(\rho)}{\SE \check{C}_E \widetilde{T}_E^2}
 P_{\CA} \left(
 \frac{1}{\SE \check{C}_E \widetilde{T}_E}
 \right)
 \nonumber
 \\
 \EQAMK + \EQAMK
 \widetilde{H}(\widetilde{T}_E-\widetilde{T}_E^{\mathrm{div}}(\rho))
 \widetilde{P}_E^\mathrm{(P)}(\widetilde{T}_E;\rho).
 }

 \begin{figure}
  \begin{center}
  \includegraphics[width=8.8cm]{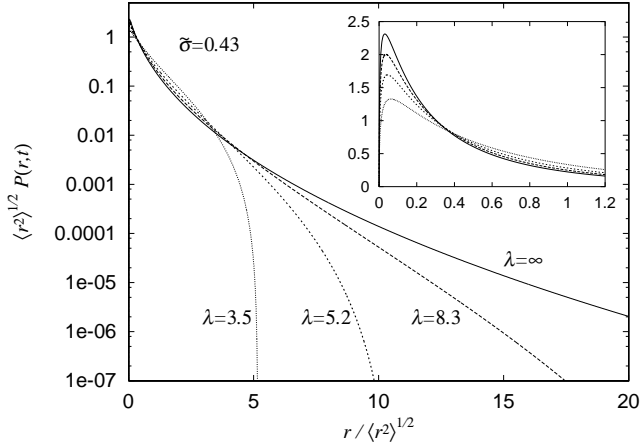}
  \end{center}
  \caption{\label{fig:TM-fPDF}
  Similarity solution (fixed-time PDF of relative separation)
  of the self-similar telegraph model in the case of
  2D-FC turbulence.
  The parameters used in this figure are estimated from
  DNS data for NV11
  by using
  the exit-time PDF.
  Different lines refer to different values of $\lambda$.
  $\lambda=\infty$ corresponds to the similarity solution of Palm's
  equation.
  The inset is a blowup of the head region in the linear scale.
  }
 \end{figure}

 These results support the self-similar telegraph model.
In the model, 
 $\widetilde{P}_E^\mathrm{(I)}(\widetilde{T}_E)$ is approximated by a
 $\delta$-function.
 On the other hand, for slowly-separating particle pairs,
 the model is approximated by Palm's equation. That is,   
 the head region of the fixed-time PDF as well as the tail of the 
 exit-time PDF is  approximately described by Palm's equation.
 By taking advantage of these characteristics of the exit-time PDF,
 we can estimate the parameters of the model, $\lambda$ and
 $\tilde{\sigma}$ ($\delta$):
 $\lambda$ is estimated in the Region-I and $\delta$ is 
 in the Region-II.
 By our DNS of 2D-FC turbulence,
 these parameters are estimated as
 $3.5\lesssim\lambda\lesssim8.3$ and $\tilde{\sigma}\approx0.43$
 ($\delta\approx-0.77$).
 The estimated value of $\tilde{\sigma}$ corresponds to the negative
 drift in the model.
 This is the opposite to the 2D-IC turbulence case \cite{OT2006b}.
 We speculate that the difference in the sign of $\widetilde{\sigma}$ is
 caused by the difference in coherent structures.
 With these parameters, we can numerically calculate
 the similarity solution of the self-similar telegraph model,
 which is shown in Fig. \ref{fig:TM-fPDF} \cite{OT2006b}.
 However we cannot compare the solution directly with the results of
 DNS because the inertial range achieved by the DNS is too narrow.
 We need DNS with higher resolutions to the comparison.

 Although the self-similar telegraph model can capture
 the essential characteristics of the relative separation 
 process,
 the distribution of the relative velocity is crucial
 to understand further details of separation processes
 and their relation to the coherent structures,
  as suggested by the exit-time PDF obtained by DNS.
 In fact, we have obtained  preliminary
 results showing that
 the distribution of the relative velocity in the inertial range
 is tightly  connected to  that in the dissipation range
 \cite{OT2006a}.
 This strongly indicates that the distribution
 directly relates to the coherent structures
 because
 their length  are of the order of  the inertial range 
 and their width  of the order  of the dissipation  range 
 \cite{TM2003}.
 An extension of  the model to include the distribution of relative velocity,
 as well as an extension to the dissipation range,
 will be done in the future work.

 \begin{acknowledgments}
  This work was supported by the Grant-in-Aid for the 21st Century COE
  ``Center for Diversity and Universality in Physics'' from the Ministry
  of Education, Culture, Sports, Science and Technology (MEXT) of Japan.
  Numerical computation in this work was carried out on a NEC SX-5
  at the Yukawa Institute Computer Facility.
 \end{acknowledgments}

 \appendix

 \section{Properties of 2D-FC turbulence}
 The governing equations of the 2D-FC turbulence are
 \EQA{
 \nabla\cdot\bm{u} = 0,
 \label{eq:incompressibility}
 \\
 \frac{\PD\bm{u}}{\PD t}
 + (\bm{u}\cdot\nabla)\bm{u}
 = -\frac{\nabla p}{\rho_0} + \nu\triangle\bm{u} - \alpha g T
 \bm{e}_g,
 \label{eq:FC2D-V}
 \\
 \frac{\PD T}{\PD t}
 + (\bm{u}\cdot\nabla)T = \kappa\triangle T,
 \label{eq:FC2D-T}
 }
 where $\bm{u}$, $T$, and $p$ represent velocity, temperature,
 and pressure field, respectively.
 $\bm{e}_g$ is the unit vector in the direction of  the
 gravity.
 $\nu$, $\kappa$, $\rho_0$, $\alpha$, and $g$ are the kinematic
 viscosity, the molecular diffusivity, the mean density of the fluid,
 the thermal expansion
 coefficient, and the gravitational acceleration, respectively.

 The 2D-FC turbulence has two important properties: the Bolgiano-Obukhov
 scaling and the fine coherent structures.

 \subsection{Bolgiano-Obukhov scaling}
 In this system, the integral of the squared temperature,
 \EQ{
 S=
 \frac{1}{L^2}
 \int \frac{T^2}{2} d\bm{x},
 }
 is a conserved quantity in the
 ideal case,
 where $L^2$ is the volume of the system.
 We call this quantity entropy for convenience.
 The entropy cascades from large to small
 scales similar to the
 energy cascade in the 3D-NS turbulence,
 which leads scaling laws of the energy and entropy spectra,
 $E(k)$ and $S(k)$:
 \EQA{
 E(k)=K_E
 {\epsilon_\theta}^{2/5}(\alpha g)^{4/5}k^{-11/5},
 \label{eq:BO-Ek}
 \\
 S(k)=K_S
 {\epsilon_\theta}^{4/5}(\alpha g)^{-2/5}k^{-7/5},
 \label{eq:BO-Sk}
 }
 where $K_E$ and $K_S$
 are considered to be universal constants, and
 $\epsilon_\theta$ is the entropy dissipation rate defined as follows:
 \EQ{
 \epsilon_\theta =
 \frac{\kappa}{L^2}
 \left\langle
 \int \nabla T\cdot\nabla T d\bm{x}
 \right\rangle.
 \label{eq:def-edrate}
 }
 In this paper, $\langle\cdot\rangle$ denotes an ensemble average.
 These scaling laws, Eqs.\ (\ref{eq:BO-Ek}) and (\ref{eq:BO-Sk}), are
 called the Bolgiano-Obukhov scaling \cite{MY1975}.

 Entropy dissipation length scale $\eta_\theta$ and time scale
 $\tau_\theta$ is estimated by dimensional analysis with
 $\epsilon_\theta$, $\alpha g$, and $\kappa$:
 \EQA{
 \eta_\theta = {\epsilon_\theta}^{-1/8}(\alpha g)^{-1/4}\kappa^{5/8},
 \label{eq:def-etatheta}
 \\
 \tau_\theta = {\epsilon_\theta}^{-1/4}(\alpha g)^{-1/2}\kappa^{1/4}.
 \label{eq:def-tautheta}
 }
 These scales correspond to the Kolmogorov scales in the 3D-NS turbulence.
 There are several other quantities characterizing 2D-FC turbulence:
 the thermal Taylor microscale $\lambda$, and the Rayleigh number at
 $\lambda$, $\mathrm{Ra}_\lambda$.
 These are defined as $\lambda=\frac{1}{2}(\lambda_x + \lambda_y)$
 and
 $\mathrm{Ra}_\lambda = \frac{1}{2} (\mathrm{Ra}_{\lambda_x} +
 \mathrm{Ra}_{\lambda_y})$, where
 \EQA{
 \lambda_x = \sqrt{
 \frac{\langle T^2\rangle}{\langle(\PD T/\PD x)^2\rangle}
 },
 \\
 \lambda_y = \sqrt{
 \frac{\langle T^2\rangle}{\langle(\PD T/\PD y)^2\rangle}
 },
 \\
 \mathrm{Ra}_{\lambda_x} =
 \frac{\alpha g\langle T^2\rangle^{1/2}\lambda_x^3}{\kappa\nu},
 \\
 \mathrm{Ra}_{\lambda_y} =
 \frac{\alpha g\langle T^2\rangle^{1/2}\lambda_y^3}{\kappa\nu}.
 }
 
 According to the Bolgiano-Obukhov scaling, Richardson's law in
 the 3D-NS turbulence is modified as follows:
 \EQA{
 \langle r(t)^p\rangle
 =
 g_p {\epsilon_\theta}^{p/2}(\alpha g)^p t^{5p/2},
 }
 where $r(t)$
 is
 relative separation of passive particles at time
 $t$ and
 $g_p$ is considered to be a universal constant.
 In addition, Richardson's diffusion equation of the separation PDF
 $P(r,t)$
 is modified as follows:
 \EQA{
   \frac{\PD P}{\PD t}
   = \frac{\PD}{\PD r} \left[
   \{k_0 \epsilon_\theta^{1/5}(\alpha g)^{2/5} r^{8/5}\}
   r\frac{\PD}{\PD r} 
   \left(\frac{P}{r}\right)
   \right],
   \label{eq:2dfc-richardson}
 }
 where $P(r,t)$ is the probability density of relative separations
 at $r$ and $t$,
 and $k_0$ is considered to be a universal
 constant.
 This equation has a self similar solution if the initial condition is
 the delta function.
 For $P(r,0)=\delta(r)$,
 \EQA{
   P(r,t) =
   \frac{2\pi r C_R}{\epsilon_\theta(\alpha g)^2(k_0 t)^5}
   \exp\left[
   -\frac{25}{4}\frac{r^{2/5}}{k_0\epsilon_\theta^{1/5}
   (\alpha	g)^{2/5} t}
   \right],
   \label{eq:2dfc-richardson-solution}
 }
 where $C_R$ is the normalization constant.

 \subsection{Fine coherent structures}
 \begin{figure*}[t]
  \begin{center}
   \includegraphics[width=8.8cm]{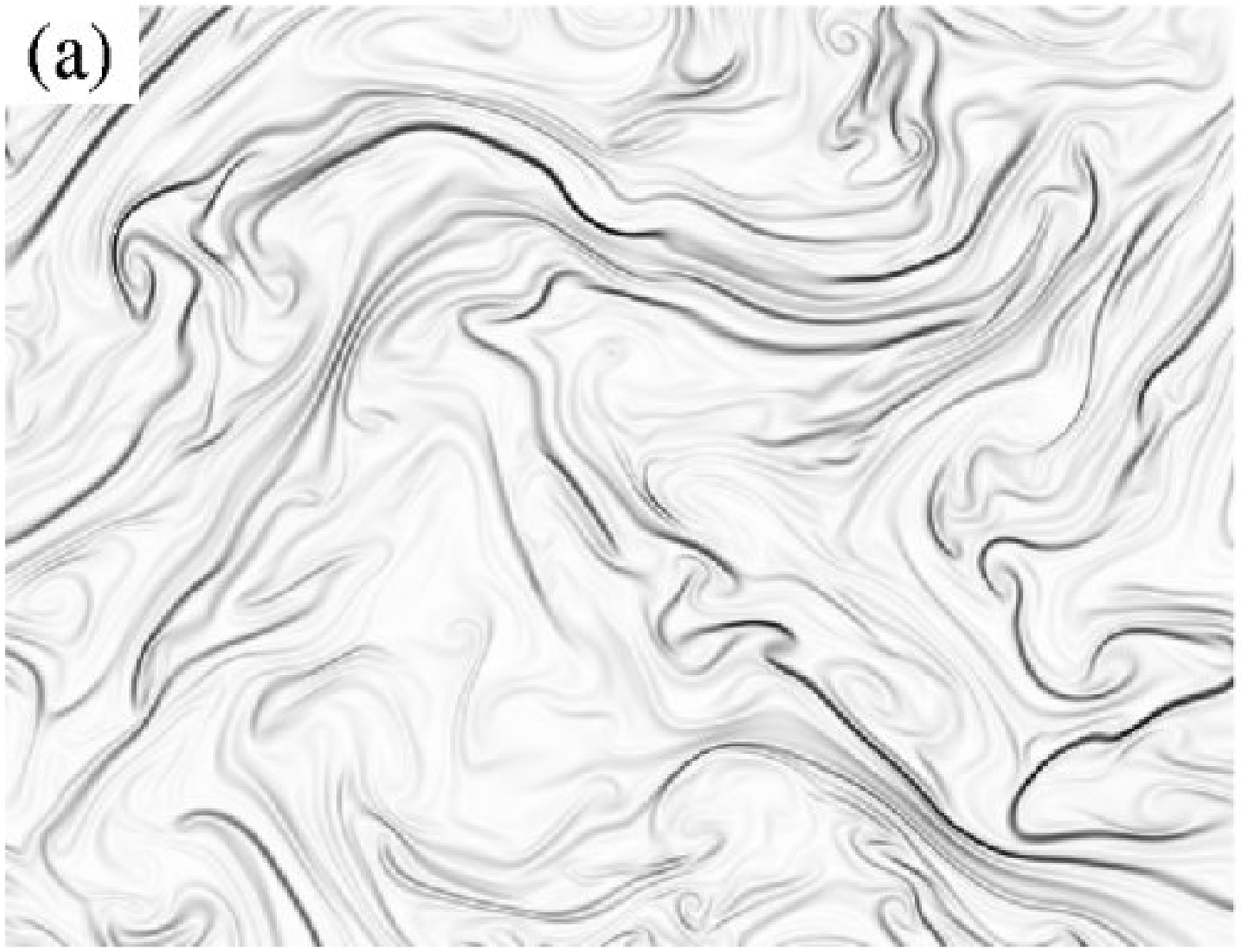}
   \includegraphics[width=8.8cm]{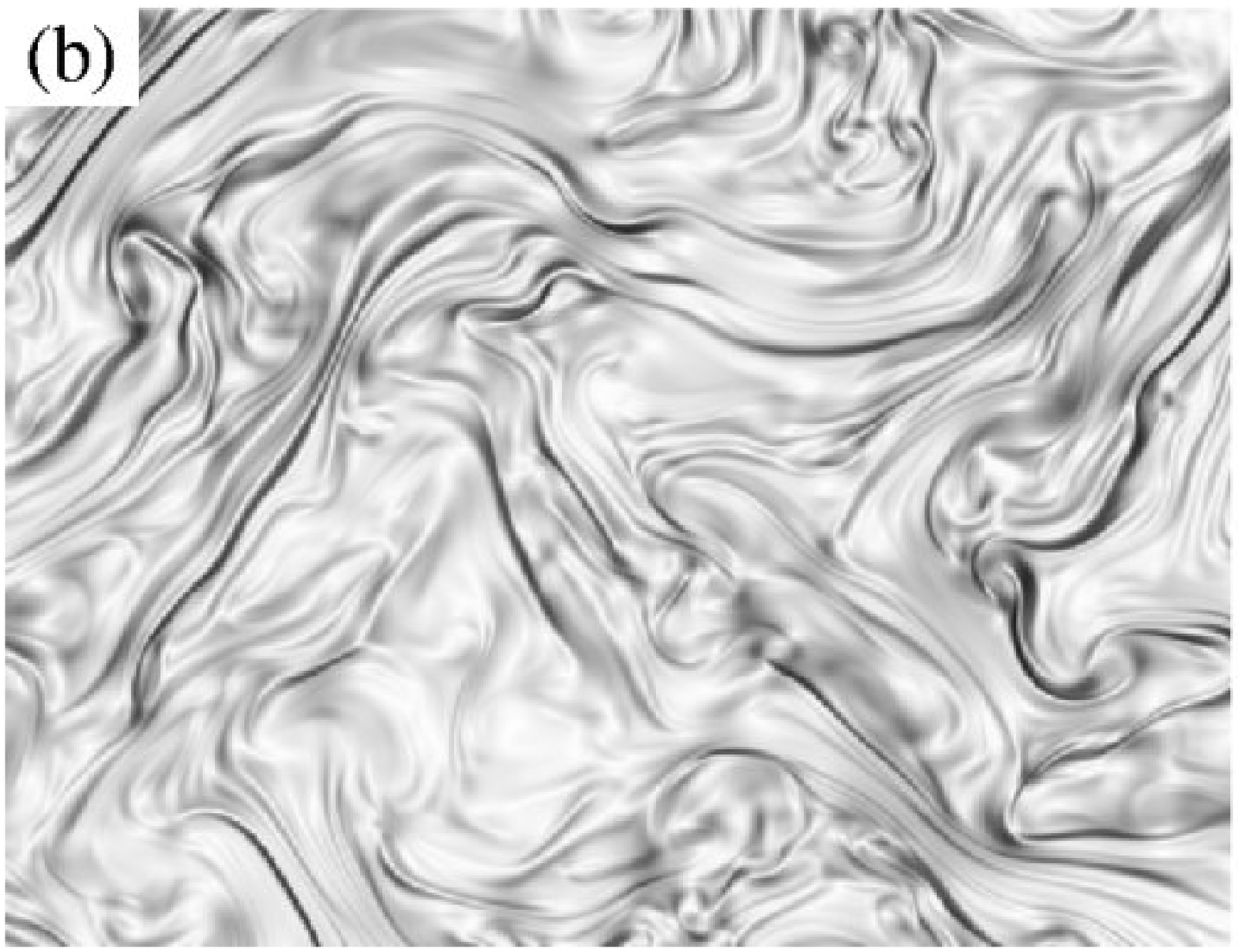}
  \end{center}
  \caption{\label{fig:snap_TV_A}
  A snapshot of T-Vorticity and strain field.
  (a) a snapshot of T-Vorticity field,
  (b) that of strain field  at the same location and time.
  Shading represents intensity of the field.
  }
 \end{figure*}

 We introduce a vector quantity called T-Vorticity,
 $\bm{\chi}(\bm{x},t)$,
 as follows:
 \EQA{
 \bm{\chi}(\bm{x},t) =
 \left(
 \frac{\PD T}{\PD y} , -\frac{\PD T}{\PD x}
 \right)
 .
 }
 
 Figure \ref{fig:snap_TV_A} shows a snapshot of the magnitude of
 the T-Vorticity and the strain field.
 It is observed that there are linearly concentrated area of
 T-vorticity.
 Its width and length are
 an order of the entropy dissipation scale and
 the integral scale, respectively \cite{TM2003}.
 We call these fine coherent structures.
 At the same location of the
 structures,
 we can observe
 linearly straining regions
 in Fig.\ \ref{fig:snap_TV_A}(b).
 
 T-Vorticity has similar properties to vorticity in the 3D-NS turbulence.
 The evolution equation of the T-Vorticity has the stretching term and is
 the same as that of vorticity in 3D-NS
 system:
 \EQA{
 \frac{\PD\bm{\chi}}{\PD t}
 + (\bm{u}\cdot\nabla)\bm{\chi}
 =
 (\bm{\chi}\cdot\nabla)\bm{u} + \kappa\triangle\bm{\chi}.
 }
 Accordingly, there is a solution corresponding to the Burgers vortex in
 the 3D-NS system, which is called the Burgers T-vortex layer.
 It is numerically confirmed that fine coherent structures in the 2D-FC
 turbulence are well approximated by the Burgers T-vortex layer
 \cite{TM2003}.
 This is similar to fine coherent structures such as worms in the 3D-NS
 turbulence which are well approximated by the Burgers vortex.

 \section{exit-time statistics}
 The exit-time statistics is one of the scale-fixed statistics.
 The exit-time, $T_\mathrm{E}(R; \rho)$, is defined as
 \EQA{
 T_\mathrm{E}(R; \rho) \equiv T_\mathrm{F}(\rho R) -
 T_\mathrm{F}(R),
 }
 where $T_\mathrm{F}(R)$ is the time when a relative separation
 $r(t)$ reach a threshold $R$ for the first time (first-passage
 time) \cite{ABCCV1997,BCCA1999,BS2002a}.
 According to the scaling law of the characteristic time, $T_c\propto
 r^\SE$,
 and the additivity of the mean,
 the form of the mean exit-time is expected as follows:
 \EQA{
 \langle T_E(R;\rho)\rangle
 =
 \check{C}_E (\rho^\SE-1)r^\SE,
 \label{eq:etime-general}
 }
 where $\check{C}_E$ is a constant depending on the system.
 In the case of the Bolgiano-Obukhov scaling,
 $\check{C}_E=C_E^\mathrm{(BO)}(\alpha g)^{-2/5}\epsilon_\theta^{-1/5}$,
 where $C_E^\mathrm{(BO)}$ is considered to be a non-dimensional
 universal constant.

 The exit-time statistics has two advantages over usual fixed-time
 statistics:
 \begin{enumerate}
  \item The exit-time statistics can specify a spatial scale by choosing
	a threshold $R$.
	Therefore we can extract information of the inertial range if
	both $R$ and $\rho R$ are in the inertial range.
  \item The interval between thresholds (the width for averaging) can be
	controlled by $\rho$.
	This means that we can control the degree of coarse graining of
	dispersion process.
 \end{enumerate}
 
 In order to calculate exit-times, we prepare a set of thresholds
 $R_n=\rho^n R_0$,
 where $n$ is a positive integer,
 and record the time at which $r(t)$ reach
 $R_n$ for the first time for every particle pairs and every $n$.
 In the present work, we set $\rho$ to $1.1^{1/8}$ $(\approx1.01)$
 and $R_0$ to the grid size $\Delta x$.
 
 If Palm's 
 equation, Eq.\ (\ref{eq:Palm-eq}),
 describes relative dispersion well,
 from the equation
 we can calculate the probability density function (PDF) of exit-time
 $P_E(T_E;R,\rho)$, which represents the probability density for
 a particle pair of which exit-time from $R$ to $\rho R$ is $T_E$,
 \cite{BS2002a}.
 In the present case, 
 to calculate the PDF,
 first we solve Eq.\ (\ref{eq:Palm-eq}) with the initial condition
 \EQA{
 P(r,0)= \delta(r-R),
 }
 where
 the boundary conditions are the reflecting condition at $r=0$ and
 the absorbing condition at $r=\rho R$.
 And then, the PDF is obtained as the time derivative of the
 probability of $r(t)<\rho R$:
 \EQA{
 P_E(T_E;R,\rho) =
 \left.
 -\frac{d}{dt}\int_{r<\rho R} P(r,t) dr
 \right|_{t=T_E}.
 \label{eq:palm-epdf}
 }
 We have numerically calculated the PDF of exit-time by
 the above procedure and compared with the PDF obtained by
 our DNS.

\end{document}